\documentclass[12pt]{article}

\usepackage[utf8]{inputenc}
\usepackage[T1]{fontenc}
\usepackage{amsmath,amssymb,bm}
\usepackage{graphicx}
\usepackage[margin=1in]{geometry}
\usepackage{xcolor}
\usepackage{booktabs}
\usepackage{array}
\usepackage{float}
\usepackage{siunitx}
\usepackage{setspace}
\usepackage[super]{cite}
\usepackage[hidelinks]{hyperref}
\graphicspath{{figures/}}
\title{A closed-form design rule for photonic hopfions of arbitrary order}
\author{\parbox{0.95\linewidth}{\centering Ali Nadarloo\textsuperscript{1}, Mohammad Javad Hajiahmadi\textsuperscript{1,*}\\ and Mohammad Soleimani\textsuperscript{1}\\[0.7em]
\small \textsuperscript{1}School of Electrical Engineering, Iran University of Science and Technology,\\ \small Tehran 1684613114, Iran\\
\small Emails: ali\_nadarloo@elec.iust.ac.ir; hajiahmadi@iust.ac.ir; soleimani@iust.ac.ir\\
\small \textsuperscript{*}Correspondence: hajiahmadi@iust.ac.ir}}
\date{}

\begin{document}
\maketitle

\begin{abstract}
Topological textures of light are robust against perturbation and are emerging as carriers of information and probes of light--matter interaction. The most intricate of them, the photonic hopfion, in which the loops of different polarization states are linked, has so far been assembled from chosen superpositions of laser modes, with weights set by hand or found by numerical optimization. Here we replace this search with a design rule. Two scalar beams, an optical vortex of order $\ell$ and a field with $n$ dark rings, propagate in closed form, and their phases on one plane are the two masks of a metasurface that controls two orthogonal polarizations. Although a beam, unlike the closed spaces of topology, extends to infinity, we prove that it is a hopfion of order $n\ell$ whenever three algebraic conditions on the ring coefficients and the vortex order hold, and that the polarization loops of every state that the far field does not reach then link exactly $n\ell$ times; each condition guards against one way in which the topology can escape. The order is thus chosen on paper and the device follows directly, without optimization. In scalar simulation, one terahertz metasurface geometry produces hopfions up to order nine, and a full-wave simulation of the complete first-order device preserves its topology. The rule applies to any platform that shapes the phases of two orthogonal polarizations.
\end{abstract}

\section*{Introduction}
A hopfion is a particle-like texture built on the Hopf map, which sends the three-sphere onto the two-sphere so that every point of the two-sphere has a circle as its preimage and any two such circles are linked once~\cite{Hopf1931}. The same structure appears as knotted solitons in field theory~\cite{Faddeev1997}, as magnetic textures in multilayers~\cite{Kent2021} and as knotted solutions of Maxwell's equations~\cite{Ranada1989,Kedia2013}. In a light beam the two-sphere is the Poincar\'e sphere of polarization states, and a photonic hopfion is a polarization field in which, in the ideal case of a texture that is uniform at infinity, every state occupies a closed loop and any two loops are linked~\cite{Sugic2021,Shen2023AP}. Photonic hopfions belong to the wider family of topological textures of light, which started with optical skyrmions~\cite{Tsesses2018} and now spans many structured fields~\cite{Shen2024NP,Shen2025AOP,He2022LSA}.

Photonic hopfions have been generated by superposing Laguerre--Gauss modes in two polarizations with spatial light modulators~\cite{Sugic2021,Shen2023AP,Ehrmanntraut2023}. The same modal approach produced higher-order hopfions~\cite{Ehrmanntraut2023}, hopfions with independently chosen poloidal and toroidal windings~\cite{Zeng2026}, space-time hopfion crystals~\cite{Lin2025PRL}, photonic spin hopfions~\cite{Wang2023PRL} and hopfion textures transferred to materials~\cite{Tamura2024}; scalar hopfions~\cite{Wan2022} and knotted polarization singularities~\cite{Larocque2018,Dennis2009} have also been reconstructed in three dimensions. Most recently, Vogliardi et al.~\cite{Vogliardi2026} encoded two circular-polarization channels in a single metasurface and measured hopfion fibrations of low order. In all of these works the order is set by the choice of superposed modes, and in the single-metasurface implementation~\cite{Vogliardi2026} the mode weights were found by numerical optimization. Metasurfaces can impose independent phases on two orthogonal polarizations~\cite{Mueller2017,Arbabi2015} and control polarization along the propagation direction~\cite{Rubin2021,Dorrah2021NP,Dorrah2021NC,Liu2024PRJ,Wang2024NC,Tan2026LSA,Zong2026LPR}; they have produced polarization knots~\cite{Intaravanne2022,Li2024ACS}, skyrmions~\cite{MataCervera2025,Fu2026,Park2026} and other topological waves~\cite{Xie2026}, but inverse design must be repeated for every target, and the invariant is checked only afterwards~\cite{Park2026}. What is missing is a rule that runs the other way: one that starts from the desired order, returns the device directly, and states in advance when the result is a true hopfion.

\begin{figure}[!htbp]
\centering
\includegraphics[width=\linewidth]{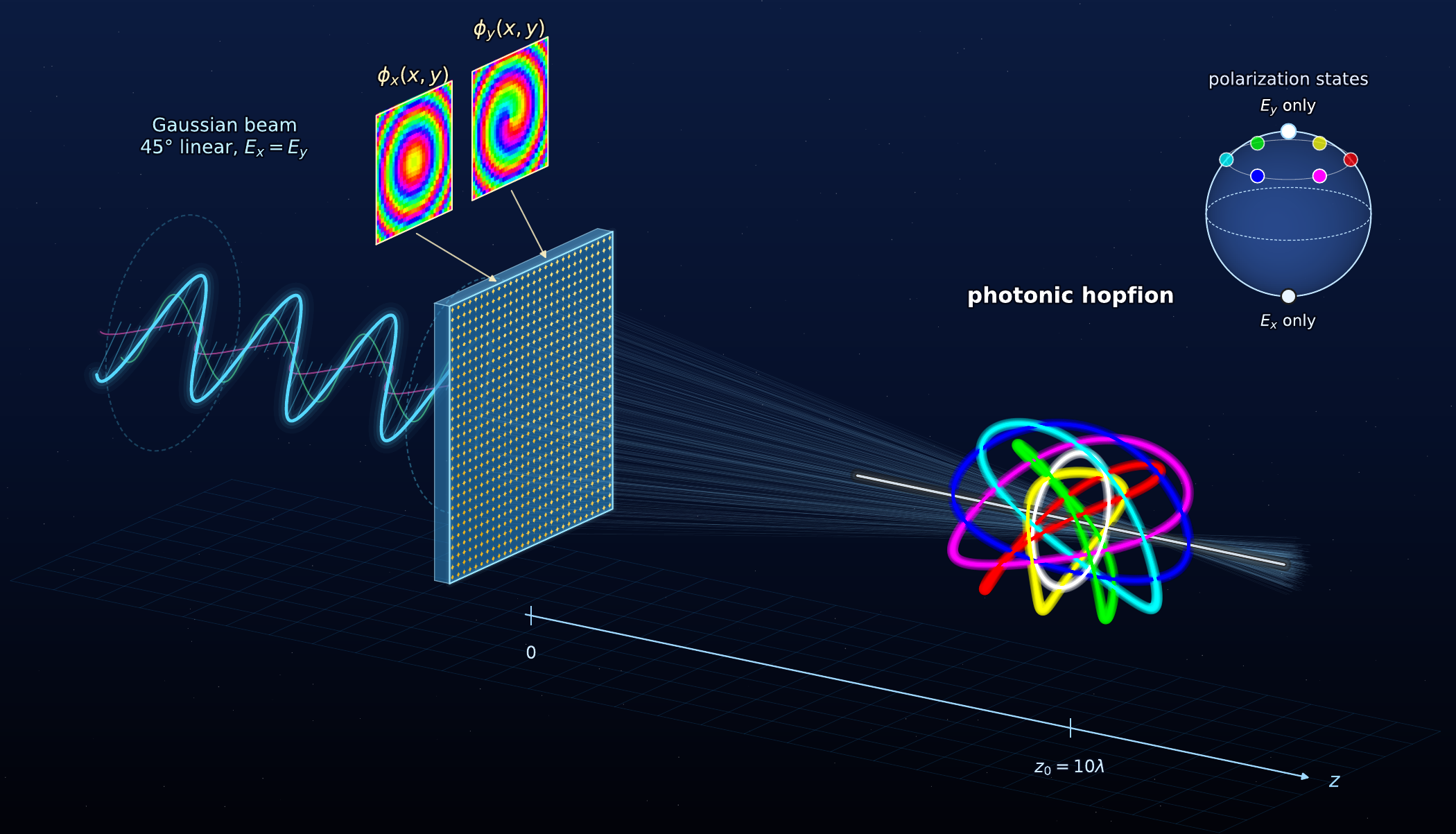}
\caption{\textbf{A photonic hopfion from one two-phase metasurface} (schematic). A Gaussian beam polarized at \ang{45} ($E_x=E_y$) illuminates a metasurface whose cells impose the phases $\phi_x(x,y)$ and $\phi_y(x,y)$ on the two linear polarizations. The transmitted beam forms a Hopf texture near $z_0=10\lambda$. The colored curves show selected closed preimages of the ideal target (states marked on the Poincar\'e sphere); the $x$-polarized state occupies the beam axis and the $y$-polarized state a ring around it (white).}
\label{fig:concept}
\end{figure}

Here we derive such a rule. We choose two scalar beams by their zeros, an axial vortex of order $\ell$ and $n$ dark rings, and propagate them in closed form; their phases at one plane are the two masks of a metasurface. We prove that the beam is a hopfion of order $n\ell$ whenever three conditions on the ring coefficients and the vortex order hold; the conditions are what make the product of the two windings survive, in a beam that is open at infinity, as the invariant of the propagating field and as the linking number of its polarization loops. The order is chosen first and the masks follow, which turns hopfion design from a search into a calculation. We demonstrate the rule with one terahertz metasurface geometry that produces hopfions up to order nine in scalar simulation, and with a full-wave simulation of the complete first-order device (Fig.~\ref{fig:concept}).

\section*{Results}
\subsection*{Two beams chosen by their zeros}
The metasurface has two local controls, the phases it imposes on $x$- and $y$-polarized light:
\begin{equation}
J(x,y)=\begin{pmatrix} e^{\mathrm{i}\phi_x(x,y)} & 0\\ 0 & e^{\mathrm{i}\phi_y(x,y)}\end{pmatrix}.
\label{eq:jones}
\end{equation}
A beam polarized at \ang{45} leaves the sheet as two independent scalar fields, $B\equiv E_x$ and $A\equiv E_y$. Their ratio $w=A/B$ fixes the local polarization, the point
\begin{equation}
\mathbf n=\frac{1}{S_0}\bigl(2\,\mathrm{Re}(A^{*}B),\;2\,\mathrm{Im}(A^{*}B),\;|A|^2-|B|^2\bigr),\qquad S_0=|A|^2+|B|^2 ,
\label{eq:stokes}
\end{equation}
on the Poincar\'e sphere. Its north pole is $y$-polarized light ($B=0$) and its south pole $x$-polarized light ($A=0$). The preimage of a state $w$, the zero set of $A-wB$, is generically a curve, and two closed preimages are linked by the Gauss integral
\begin{equation}
Lk(L_1,L_2)=\frac{1}{4\pi}\oint_{L_1}\!\oint_{L_2}\frac{(\mathbf r_1-\mathbf r_2)\cdot(\mathrm{d}\mathbf r_1\times\mathrm{d}\mathbf r_2)}{|\mathbf r_1-\mathbf r_2|^3}.
\label{eq:gauss}
\end{equation}
The design task is therefore geometric: the sheet has to launch two waves whose zeros form a linked skeleton and whose ratio fills the sphere of states between them.

We prescribe both components in a focal plane $z=z_0$ behind the sheet. With polar coordinates $(r,\theta)$ and $\rho=r/w_0$,
\begin{equation}
A_f=\rho^{\ell}e^{\mathrm{i}\ell\theta}e^{-\rho^2},\qquad B_f=\sum_{p=0}^{n}c_p\,L_p(2\rho^2)\,e^{-\rho^2},
\label{eq:target}
\end{equation}
where $L_p$ are the Laguerre polynomials and the real coefficients $c_p$, with $c_n\neq0$, are the only design parameters. $A_f$ is a vortex of order $\ell$~\cite{Allen1992}, dark only on the axis. $B_f$ is a superposition of radial modes, dark on rings at the positive roots of $P(x)=\sum_pc_pL_p(2x)$, $x=\rho^2$; its simplest member, a single ring, is the seed of the first photonic hopfions~\cite{Sugic2021,Shen2023AP}. Both components propagate in closed form (Supplementary Note~1). With $z_R=kw_0^2/2$, $\nu=1+\mathrm{i}(z-z_0)/z_R$, the Gouy angle $\psi=\arg\nu$ and the co-moving radius $\xi=r/(w_0|\nu|)$~\cite{Siegman1986},
\begin{equation}
\frac{A}{g}=\xi^{\ell}e^{\mathrm{i}\ell(\theta-\psi)},\qquad \frac{B}{g}=\sum_{p=0}^{n}c_p\,L_p(2\xi^2)\,e^{-2\mathrm{i} p\psi},
\label{eq:lift}
\end{equation}
where the Gaussian beam $g=\nu^{-1}e^{-\rho^2/\nu}$ never vanishes and drops out of the polarization. Each radial mode keeps its shape in $\xi$ and acquires its own Gouy phase. In these coordinates all of space becomes the slab $|\psi|<\pi/2$, whose faces $\psi=\pm\pi/2$ are the far field, where $B/g$ tends to
\begin{equation}
F(\xi^2)=\sum_{p=0}^{n}(-1)^p\,c_p\,L_p(2\xi^2).
\label{eq:far}
\end{equation}

\subsection*{Why the order is $n\ell$}
\begin{figure}[!htbp]
\centering
\includegraphics[width=\linewidth]{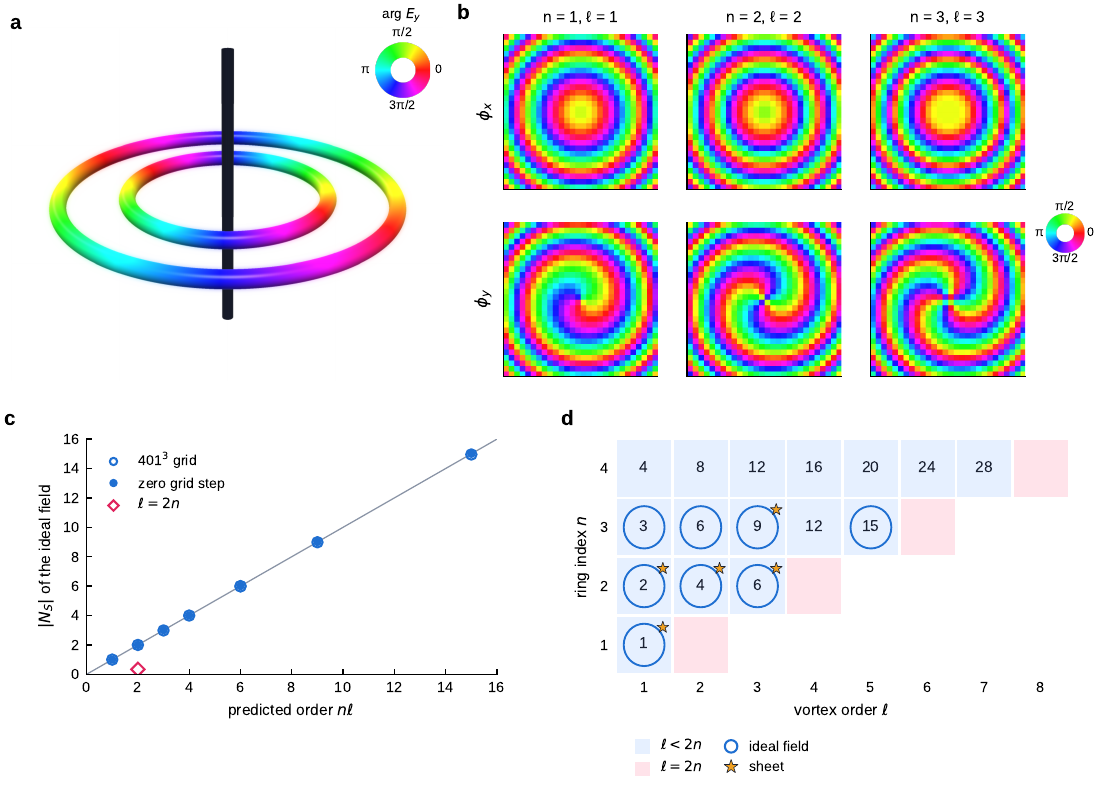}
\caption{\textbf{The design rule.} \textbf{a},~Zero-set skeleton of the ideal field for $n=2$, $\ell=2$: the axis, where $E_y=0$, and the two dark rings, where $E_x=0$, colored by $\arg E_y$, which turns $\ell$ times along each ring. Coordinates $x/w_0$, $y/w_0$ and $(z-z_0)/z_R$. \textbf{b},~Phase maps of the $28\times28$ sheet, equation~\eqref{eq:phases}, for $(n,\ell)=(1,1)$, $(2,2)$ and $(3,3)$, with the ring components $(c_0,c_1)=(\tfrac34,\tfrac14)$, $(3,2,2)$ and $(4,3,2,2)$. \textbf{c},~Skyrme number of the ideal fields over all space versus the predicted order $n\ell$ on the finest grid (open circles) and extrapolated to zero grid step (dots); diamond, the boundary case $\ell=2n$ of the single ring. \textbf{d},~Order chart: every entry $(n,\ell)$ with $\ell<2n$ is a hopfion of order $n\ell$. Circles, ideal fields evaluated in \textbf{c}; stars, order of the designed rings of the sheets in Fig.~\ref{fig:orders}.}
\label{fig:rule}
\end{figure}

\begin{figure}[!htbp]
\centering
\includegraphics[width=\linewidth]{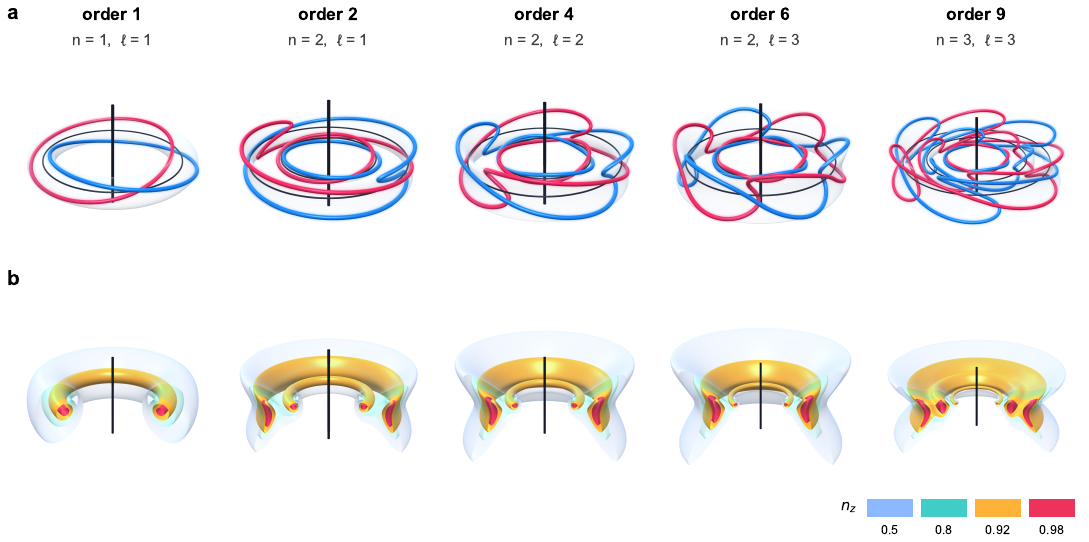}
\caption{\textbf{Ideal hopfions of order 1 to 9 from the closed form.} \textbf{a},~Fibers of two polarization states, $\arg(E_y/E_x)=0$ (red) and $\pi$ (blue), on the tori $n_z=0.88$ around each dark ring (gray), with the axis $E_y=0$ (black). Each fiber winds $\ell$ times around its ring, so the two states link $n\ell$ times. \textbf{b},~Cut-away view of the nested surfaces of constant $n_z$. Coordinates as in Fig.~\ref{fig:rule}a; the components are scaled to equal power in the focal plane.}
\label{fig:hopfions}
\end{figure}

The zeros already show the order (Fig.~\ref{fig:rule}a). The axis is $x$-polarized, the dark rings are $y$-polarized, and along every ring the phase of $A$ turns $\ell$ times, because the axis is a vortex of order $\ell$. Near a ring the surfaces of constant $n_z$ close into thin tori, and on such a torus the loop of one polarization state winds $\ell$ times around the ring while it goes once around the axis. Two states are therefore linked $\ell$ times per ring, and $n$ rings of equal orientation give $n\ell$ (Fig.~\ref{fig:hopfions}). The picture leaves two questions open: do all rings have the same orientation, and does any part of the structure escape to infinity, where a beam, unlike a closed space, is open? The theorem answers both.

\subsection*{The order theorem}
The integer that measures a hopfion is the Skyrme number~\cite{Skyrme1961,Sugic2021}. Because the common Gaussian factor $g$ is nonzero and does not affect the polarization texture, we define the reduced fields $\hat A=A/g$ and $\hat B=B/g$. Their normalized pair
\begin{equation*}
\mathbf u=\frac{(\mathrm{Re}\hat A,\ \mathrm{Im}\hat A,\ \mathrm{Re}\hat B,\ \mathrm{Im}\hat B)}{\sqrt{|\hat A|^2+|\hat B|^2}}
\end{equation*}
is a point on the three-sphere $S^3$, and
\begin{equation}
N_S=\frac{1}{2\pi^2}\int\det\bigl[\mathbf u,\ \partial_1\mathbf u,\ \partial_2\mathbf u,\ \partial_3\mathbf u\bigr]\,\mathrm{d}^3q
\label{eq:NS}
\end{equation}
is the signed volume of $S^3$ that $\mathbf u$ covers. For a map from a closed space, $N_S$ is the degree and equals the Hopf invariant, the linking number of two preimages~\cite{Sugic2021}. Our main result states when this degree survives in a beam.

\emph{Theorem.} If (i) $F(x)>0$ for all $x\ge0$, (ii) the polynomial $\sum_pc_pt^p$ has no roots with $|t|\le1$, and (iii) $\ell<2n$, then
\begin{equation}
|N_S|=n\ell .
\label{eq:order}
\end{equation}

The idea of the proof is simple. $N_S$ counts, with signs, how many times the beam reaches one chosen point of $S^3$, and this count can change only when that point is reached on the boundary of space. Adding the far-field faces and the limit $\xi\to\infty$ closes the slab into a solid cylinder. Condition (ii) keeps $B$ nonzero on the axis, condition (i) keeps it real and positive in the far field, and condition (iii) makes it dominate far from the axis, so the boundary is mapped into a disc of states around the $x$-polarized pole and a circle, a set of zero volume that does not divide $S^3$. The count can therefore be taken at a $y$-polarized state, whose preimages sit on the dark rings: each ring contributes $\ell$ times its index, and the argument principle on the boundary of the meridional half-plane gives the sum of the indices as $-n$. Hence $|N_S|=n\ell$ (Supplementary Note~1 gives the full proof~\cite{Milnor1965,Hatcher2002}).

The conditions are sufficient, not necessary, and each one guards a step of the proof. Condition (i) really only asks $F$ to keep one sign: reversing every $c_p$ flips $B$, a half-turn of the Poincar\'e sphere that changes nothing topological. A zero of $F$ is another matter, since a dark ring then reaches infinity. If (ii) fails, a ring touches the axis, or roots inside the unit circle change the count; at $\ell=2n$ the boundary divides $S^3$ and $N_S$ need not be an integer. A corollary connects the degree to the loops themselves. The far field reaches only the states with $|w|\le R_{\max}=\max_{x\ge0}x^{\ell/2}/F(x)$; every state beyond this cap has a preimage made of closed loops, and the preimages of any two such states link exactly $n\ell$ times (Supplementary Note~1). For every design of this work $R_{\max}<2.3$, so all states near the $y$-polarized pole close and link. Evaluated over all space, the Skyrme number of every design converges to $n\ell$, up to order 15 (Fig.~\ref{fig:rule}c).

\subsection*{From the field to the masks}
Because the target propagates in closed form, the masks need no inverse calculation. The sheet sits a distance $z_0$ before the focus, at $\psi=-\arctan\zeta$ with $\zeta=z_0/z_R$, and its two phases are those of $A$ and $B$ there:
\begin{equation}
\begin{aligned}
\phi_y(x,y)&=\ell\theta+(\ell+1)\arctan\zeta-\frac{\zeta}{1+\zeta^2}\,\rho^2,\\[2pt]
\phi_x(x,y)&=\arg\Bigl[\sum_{p=0}^{n}c_p\,L_p\Bigl(\frac{2\rho^2}{1+\zeta^2}\Bigr)e^{2\mathrm{i} p\arctan\zeta}\Bigr]+\arctan\zeta-\frac{\zeta}{1+\zeta^2}\,\rho^2 .
\end{aligned}
\label{eq:phases}
\end{equation}
Both masks share a converging lens of focal length $z_0(1+\zeta^{-2})$; the $y$ mask adds the spiral that creates the vortex, and the $x$ mask adds the radial phase that places the rings and fixes the order. Sampling equation~\eqref{eq:phases} at the cell centers gives the device (Fig.~\ref{fig:rule}b). A phase-only sheet transmits the incident amplitude rather than the target amplitude, so the realized beam is a deformation of the target, and topology is what survives deformation.

\subsection*{A terahertz metasurface}
\begin{figure}[!htbp]
\centering
\includegraphics[width=\linewidth]{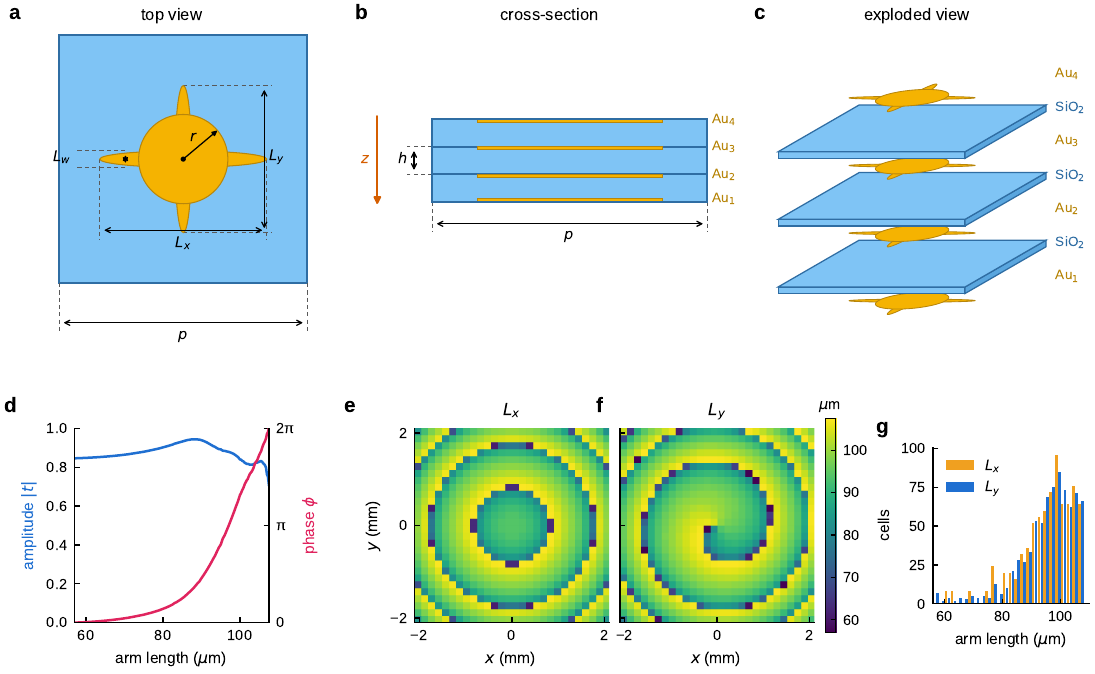}
\caption{\textbf{Two-phase meta-atom and the cells of the device.} \textbf{a},~Top view of one gold layer on the $p=\SI{150}{\micro\meter}$ lattice: a central disk ($r=\SI{27}{\micro\meter}$) and pointed elliptical arms of width $L_w=\SI{10}{\micro\meter}$ and lengths $L_x$, $L_y$. \textbf{b},~Cross-section: four \SI{0.3}{\micro\meter} gold planes separated by three fused-quartz spacers of thickness $h=\SI{15}{\micro\meter}$; light propagates along $+z$, and the exit face is the $z$ reference. \textbf{c},~Exploded view of one cell; the four gold patterns of a cell are identical. \textbf{d},~Transmission amplitude and phase at \SI{1}{\tera\hertz} versus arm length, with the orthogonal arm at \SI{85}{\micro\meter}; by symmetry the same curves hold for $L_x$ and $L_y$. \textbf{e},\,\textbf{f},~Arm lengths $L_x$ and $L_y$ selected for the $28\times28$ cells. \textbf{g},~Distribution of the selected arm lengths.}
\label{fig:unitcell}
\end{figure}

We implement the rule at \SI{1}{\tera\hertz} ($\lambda=\SI{300}{\micro\meter}$) with $w_0=1.5\lambda$ and $z_0=10\lambda$, on $28\times28$ cells of pitch $\lambda/2$, illuminated by a \ang{45}-polarized Gaussian beam. The meta-atom follows our earlier polarization-multiplexed design~\cite{Nadarloo2026}: four gold patterns in fused quartz, each a central disk with two orthogonal pairs of arms whose lengths set the two phases independently (Fig.~\ref{fig:unitcell}). The library covers \ang{360} in both channels, and the selected cells reproduce the masks with an r.m.s.\ phase error of \ang{1.8} (Supplementary Note~2).

\subsection*{Hopfions of order one to nine}
\begin{figure}[!htbp]
\centering
\includegraphics[width=\linewidth]{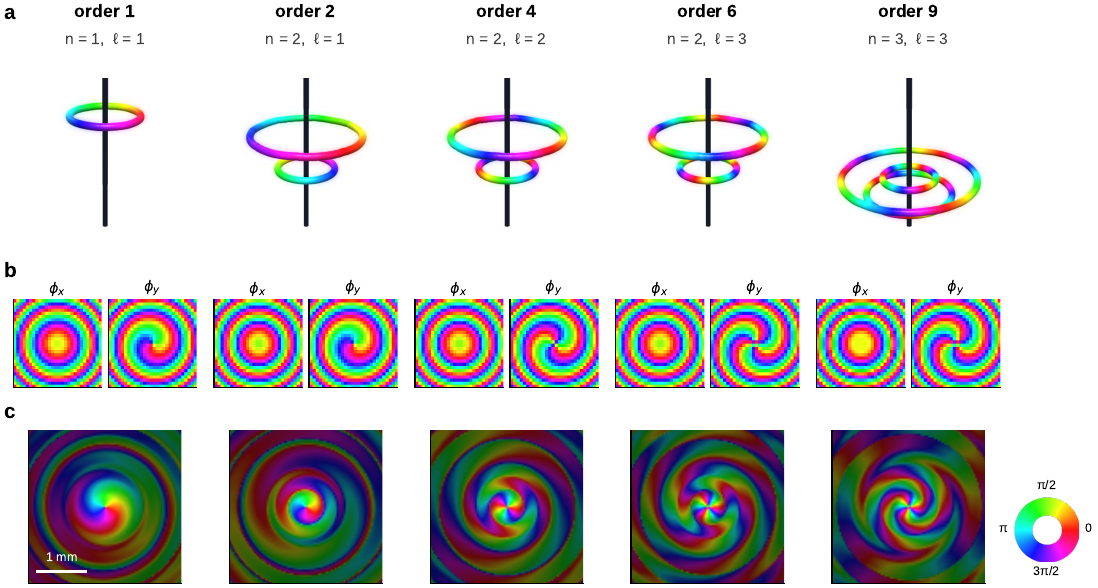}
\caption{\textbf{Hopfions of orders 1, 2, 4, 6 and 9 from the same $28\times28$ sheet geometry} (scalar model; one pair of masks per order; $(n,\ell)=(1,1)$, $(2,1)$, $(2,2)$, $(2,3)$ and $(3,3)$). \textbf{a},~Dark rings ($E_x=0$) traced in the field of the sheet, colored by $\arg E_y$, and the beam axis; each ring carries $\ell$ turns of the phase. \textbf{b},~The two phase maps. \textbf{c},~Transverse texture in the focal plane: hue, $\arg(E_y/E_x)$; brightness, transverse intensity; scale bar, \SI{1}{\milli\meter}.}
\label{fig:orders}
\end{figure}

Changing only the coefficients and $\ell$ in equation~\eqref{eq:phases} turns the same sheet geometry into hopfions of higher order; each order has its own pair of masks. We propagate each pair with the angular spectrum, including the finite aperture, the pixelation and the phase-only projection. Every sheet forms the predicted skeleton: two rings for the $(3,2,2)$ masks and three for $(4,3,2,2)$, all with the same orientation and with $\ell$ turns of the phase of $E_y$ along each (Fig.~\ref{fig:orders}a).

A finite aperture also produces weak diffraction rings in the beam periphery, so we distinguish these from the rings inherited from the ideal target. We do this without an intensity threshold by continuously morphing the ideal field into the realized field and following each target ring throughout the deformation (Methods). Every target ring survives with its orientation and its $\ell$ turns intact, so that the designed rings carry the prescribed order $n\ell$ in all seven designs, spanning orders 1 to 9 (Table~\ref{tab:summary}). The closed polarization preimages that wrap them link exactly $n\ell$ times. Additional rings can enter from the dim periphery during the morph, at no more than 0.4\% of the peak intensity; these arise from finite-aperture diffraction, are not continuations of the target rings, and add to the count taken over the whole analysis volume (Supplementary Note~4). The practical limit at higher order is instead set by the Gaussian envelope: each additional designed ring lies farther from the beam center, where the field is weaker, while increasing $\ell$ shifts more weight outward. This is why $\ell=3$ gives the brightest three-ring hopfion (Supplementary Table~1). For the three-ring designs with $\ell=1$ and 2, the outer target ring is even dimmer than some diffraction rings, but the continuation still identifies it unambiguously (Supplementary Note~4).

\subsection*{The complete device in full-wave simulation}
\begin{figure}[!htbp]
\centering
\includegraphics[width=\linewidth]{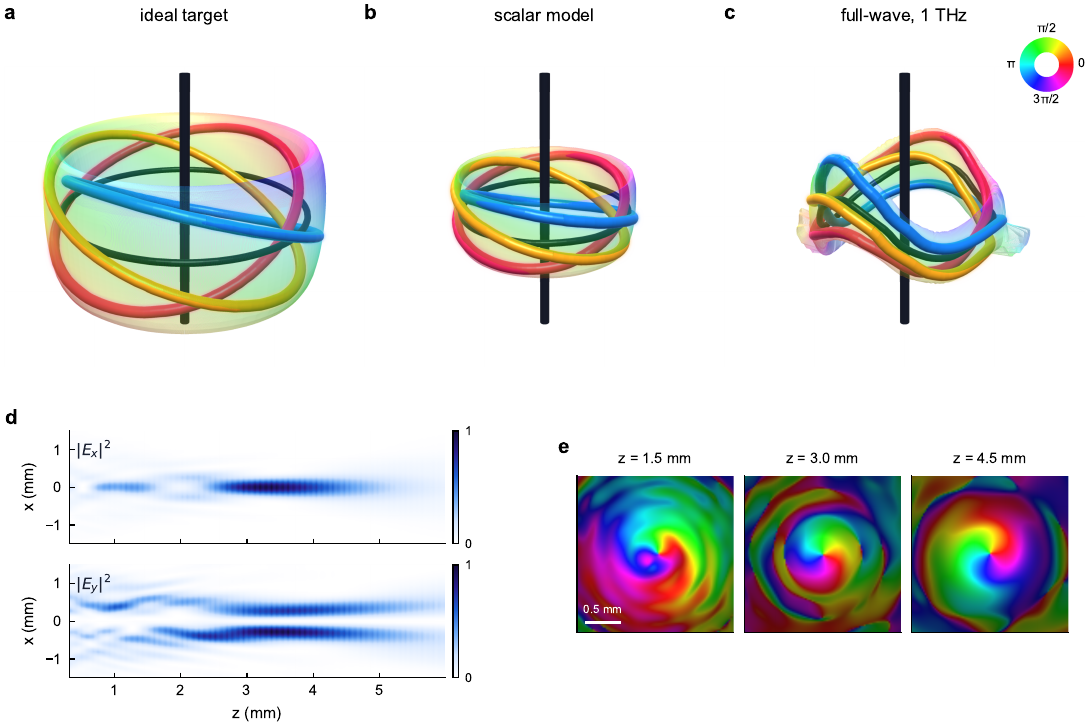}
\caption{\textbf{The first-order hopfion in the ideal target, the scalar model and the full-wave field.} \textbf{a}--\textbf{c},~Torus $n_z=0.95$ colored by $\arg(E_y/E_x)$, the dark ring $E_x=0$ (black), the preimages of $\tilde A/\tilde B=8$ (red), $8\mathrm{i}$ (gold) and $-8$ (blue), and the axis, for the ideal target (\textbf{a}), the scalar model of the sheet (\textbf{b}) and the full-wave simulation of the complete device at \SI{1}{\tera\hertz} (\textbf{c}); coordinates in millimeters. \textbf{d},~Full-wave $|E_x|^2$ and $|E_y|^2$ in the plane $y=0$, each normalized to its maximum. \textbf{e},~Full-wave transverse texture at three distances from the sheet, displayed as in Fig.~\ref{fig:orders}c.}
\label{fig:fullwave}
\end{figure}

The most demanding test keeps every effect that the scalar model omits: the vector response of each cell, the coupling between neighboring cells of different geometry, cross-polarized scattering, the longitudinal field and reflection. We simulate the complete first-order device with all 784 gold--quartz cells in place (Methods). The hopfion survives intact (Fig.~\ref{fig:fullwave}). The dark ring and the preimages of three polarization states close, their six mutual Gauss linking numbers equal one to within $10^{-3}$ (Table~\ref{tab:summary}), and the morph from the target carries the designed ring into the full-wave ring with order one. This holds although the full-wave field departs visibly from the scalar model: the topology is fixed, while the waveform is free to deform. The hopfion persists at \SI{0.975}{\tera\hertz} and begins to degrade at \SI{1.025}{\tera\hertz}, close to a resonance of the cells, and at the design frequency the sheet transmits 65\% of the incident power, equally in both channels (Supplementary Note~3).

\begin{table}[!htbp]
\centering
\caption{\textbf{Realized hopfions.} Rings, designed rings that continue the rings of the target, traced with no intensity floor; order, signed sum of the windings of $\arg E_y$ along them; $Lk$, linking of the closed preimages that wrap the designed rings ($\tilde A/\tilde B=\pm30$; full-wave, range of the six linking numbers of the ring and the preimages of $\tilde A/\tilde B=8$, $8\mathrm{i}$, $-8$); $F$, overlap with the ideal target (scalar) or with the scalar model (full-wave). $\tilde A/\tilde B$ labels states of the field scaled to equal power in both components.}
\label{tab:summary}
\small
\begin{tabular}{lccccc}
\toprule
Field & $n\ell$ & rings & order & $Lk$ & $F$\\
\midrule
Scalar sheet, $(1,1)$ & 1 & 1 & 1 & 1.000 & 0.79\\
Scalar sheet, $(2,1)$ & 2 & 2 & 2 & 2.000 & 0.88\\
Scalar sheet, $(2,2)$ & 4 & 2 & 4 & 4.000 & 0.91\\
Scalar sheet, $(2,3)$ & 6 & 2 & 6 & 6.001 & 0.91\\
Scalar sheet, $(3,1)$ & 3 & 3 & 3 & 3.000 & 0.86\\
Scalar sheet, $(3,2)$ & 6 & 3 & 6 & 6.000 & 0.89\\
Scalar sheet, $(3,3)$ & 9 & 3 & 9 & 9.000 & 0.89\\
Full-wave, $(n,\ell)=(1,1)$, \SI{1}{\tera\hertz} & 1 & 1 & 1 & 1.000--1.001 & 0.88\\
\bottomrule
\end{tabular}
\end{table}

\section*{Discussion}
The central result is a map from a topological order to a device. In practice it is a recipe: choose the order as a product $n\ell$, pick a ring polynomial of degree $n$ that passes three algebraic tests, and evaluate one formula at the cell centers. The order is known before any field is computed, no inverse optimization is needed at any order or wavelength, and the ways the guarantee can break have names: a ring reaching infinity, a ring touching the axis, or a vortex that outgrows the rings.

For maps of a closed space, a generalized Hopf map whose two complex components wind $m$ and $n$ times has degree $mn$~\cite{Milnor1965,Hatcher2002}, and the order of modal hopfions has been reported as the product of their windings~\cite{Zeng2026}. In a beam this product is not guaranteed: at $\ell=2n$ the Skyrme number is not even an integer. The theorem gives the conditions under which the propagating field keeps the degree $n\ell$, its corollary shows that the polarization loops then link $n\ell$ times, and the phase formula turns that field directly into masks, in place of the optimized mode weights of earlier metasurface hopfions~\cite{Vogliardi2026}.

The construction depends only on $c_p$, $\ell$, $\zeta$ and $w_0/\lambda$, and $A$ and $B$ may be any two orthogonal polarizations, so any platform with independent phase control of two states can implement it, from spin-decoupled metasurfaces~\cite{Vogliardi2026} to dielectric meta-atoms at optical wavelengths~\cite{Mueller2017,Arbabi2015}. The results presented here are computational; Supplementary Note~2 gives a fabrication route for the gold--quartz stack, and Supplementary Note~5 the range of focal distances, sheet sizes and fabrication errors over which the hopfion survives. Shaping the illumination beyond a single Gaussian is a natural route to still higher orders. A passive element that produces hopfions of a chosen order opens the way to studying the transport of volumetric optical topology, its transfer to matter~\cite{Tamura2024} and its robustness in complex media~\cite{Wang2024LSA}, in spectral ranges where programmable wavefront shaping is difficult.

\section*{Materials and methods}
\subsection*{Scalar propagation}
The ring components are $(c_0,c_1)=(\tfrac34,\tfrac14)$, $(3,2,2)$ and $(4,3,2,2)$; all satisfy conditions (i) and (ii), and $\ell$ is chosen within (iii). The aperture fields $u_{\rm in}e^{\mathrm{i}\phi_{x,y}}$, with $u_{\rm in}$ the illumination, are sampled on the cells and propagated with $\exp(\mathrm{i} k_zz)$, keeping only propagating waves~\cite{Goodman2005}. The overlap of two fields is
\begin{equation}
F=\frac{|\langle\mathbf E_{\rm t}|\mathbf E\rangle|^2}{\langle\mathbf E_{\rm t}|\mathbf E_{\rm t}\rangle\,\langle\mathbf E|\mathbf E\rangle},\qquad \langle\mathbf E_1|\mathbf E_2\rangle=\int_V\bigl(E_{1x}^{*}E_{2x}+E_{1y}^{*}E_{2y}\bigr)\,\mathrm{d}^3r ,
\label{eq:overlap}
\end{equation}
with both components of the target scaled to unit power. All parameters are listed in Supplementary Note~3.

\subsection*{Skyrme number of the ideal fields}
With both components scaled to equal focal power, equation~\eqref{eq:lift} is sampled over all space on a grid of up to $401^3$ points, uniform in $\psi\in[-\pi/2,\pi/2]$ and in $a_{x,y}\in(-1,1)$, where the Cartesian co-moving coordinates are $\xi_{x,y}=(x,y)/(w_0|\nu|)=\tan(\pi a_{x,y}/2)$. Each cube is divided into six tetrahedra~\cite{Freudenthal1942}, whose signed volume on $S^3$ is approximated by $\det[\mathbf u_0,\mathbf u_1,\mathbf u_2,\mathbf u_3]/6$; the error, proportional to the squared grid step, is removed by extrapolation.

\subsection*{Loops, rings and the morph}
Zeros of $B$ and of $\tilde A-w\tilde B$, with $(\tilde A,\tilde B)=(A,\kappa_{AB}B)$ scaled to equal power, are located in 120 meridional half-planes from the phase winding around each plaquette and followed from plane to plane; the linking of two closed loops is the Gauss integral, equation~\eqref{eq:gauss}. An independent grid tracer follows every zero line through the cell faces of the sampled volume. For the morph, the target is scaled by one complex factor per component and one common phase factor per plane, which leave its zeros and its topology unchanged, and every dark ring of $E_s=(1-s)E_{\rm target}+sE_{\rm realized}$ is traced without an intensity floor and matched to the nearest ring of the same orientation at the previous step. Rings reached from rings of the target form the designed structure. For the full-wave field the path runs through the scalar model.

\subsection*{Full-wave simulation}
The device was simulated with the time-domain solver of CST Studio Suite with open boundaries, lossy gold and low-loss fused quartz. The illumination was imported as a field source before the sheet, and the electric field was exported at 0.975, 1.000 and \SI{1.025}{\tera\hertz}. The analysis starts one wavelength behind the exit face.

\subsection*{Code availability}
The Python code for the design, the propagation, the topological analysis and the figures, together with its numerical outputs and the CST excitation files, is available on Zenodo at \url{https://doi.org/10.5281/zenodo.23929631}.

\subsection*{Data availability}
The full-wave field data are available from the corresponding author upon reasonable request.

\section*{Author contributions}
A.N.\ conceived the idea, derived the theory, designed the metasurface, carried out the simulations and wrote the manuscript. M.J.H.\ advised the project and contributed to the technical discussion and revision. M.S.\ supervised the work. All authors reviewed the manuscript.

\section*{Conflict of interest}
The authors declare no competing interests.

\section*{Supplementary information}
Supplementary information accompanies the manuscript.


\clearpage
\begin{thebibliography}{99}

\bibitem{Hopf1931}
Hopf, H. {\"U}ber die Abbildungen der dreidimensionalen Sph{\"a}re auf die Kugelfl{\"a}che. \emph{Math. Ann.} \textbf{104}, 637--665 (1931).

\bibitem{Faddeev1997}
Faddeev, L. \& Niemi, A. J. Stable knot-like structures in classical field theory. \emph{Nature} \textbf{387}, 58--61 (1997).

\bibitem{Kent2021}
Kent, N. et al. Creation and observation of Hopfions in magnetic multilayer systems. \emph{Nat. Commun.} \textbf{12}, 1562 (2021).

\bibitem{Ranada1989}
Ra{\~n}ada, A. F. A topological theory of the electromagnetic field. \emph{Lett. Math. Phys.} \textbf{18}, 97--106 (1989).

\bibitem{Kedia2013}
Kedia, H., Bialynicki-Birula, I., Peralta-Salas, D. \& Irvine, W. T. M. Tying knots in light fields. \emph{Phys. Rev. Lett.} \textbf{111}, 150404 (2013).

\bibitem{Sugic2021}
Sugic, D. et al. Particle-like topologies in light. \emph{Nat. Commun.} \textbf{12}, 6785 (2021).

\bibitem{Shen2023AP}
Shen, Y. et al. Topological transformation and free-space transport of photonic hopfions. \emph{Adv. Photon.} \textbf{5}, 015001 (2023).

\bibitem{Tsesses2018}
Tsesses, S. et al. Optical skyrmion lattice in evanescent electromagnetic fields. \emph{Science} \textbf{361}, 993--996 (2018).

\bibitem{Shen2024NP}
Shen, Y. et al. Optical skyrmions and other topological quasiparticles of light. \emph{Nat. Photon.} \textbf{18}, 15--25 (2024).

\bibitem{Shen2025AOP}
Shen, Y., Wang, H. \& Fan, S. Free-space topological optical textures: tutorial. \emph{Adv. Opt. Photon.} \textbf{17}, 295--374 (2025).

\bibitem{He2022LSA}
He, C., Shen, Y. \& Forbes, A. Towards higher-dimensional structured light. \emph{Light Sci. Appl.} \textbf{11}, 205 (2022).

\bibitem{Ehrmanntraut2023}
Ehrmanntraut, D. et al. Optical second-order skyrmionic hopfion. \emph{Optica} \textbf{10}, 725--731 (2023).

\bibitem{Zeng2026}
Zeng, X. et al. Optical hopfions with arbitrary two winding numbers. Preprint at https://arxiv.org/abs/2604.21424 (2026).

\bibitem{Lin2025PRL}
Lin, W., Mata-Cervera, N., Ota, Y., Shen, Y. \& Iwamoto, S. Space-time optical hopfion crystals. \emph{Phys. Rev. Lett.} \textbf{135}, 083801 (2025).

\bibitem{Wang2023PRL}
Wang, H. \& Fan, S. Photonic spin hopfions and monopole loops. \emph{Phys. Rev. Lett.} \textbf{131}, 263801 (2023).

\bibitem{Tamura2024}
Tamura, R., Allam, S. R., Litchinitser, N. M. \& Omatsu, T. Three-dimensional projection of optical hopfion textures in a material. \emph{ACS Photonics} \textbf{11}, 4958--4965 (2024).

\bibitem{Wan2022}
Wan, C. et al. Scalar optical hopfions. \emph{eLight} \textbf{2}, 22 (2022).

\bibitem{Larocque2018}
Larocque, H. et al. Reconstructing the topology of optical polarization knots. \emph{Nat. Phys.} \textbf{14}, 1079--1082 (2018).

\bibitem{Dennis2009}
Dennis, M. R., O'Holleran, K. \& Padgett, M. J. Singular optics: optical vortices and polarization singularities. \emph{Prog. Opt.} \textbf{53}, 293--363 (2009).

\bibitem{Vogliardi2026}
Vogliardi, A. et al. Optical hopfions from a single metasurface. \emph{APL Photonics} \textbf{11}, 101303 (2026).

\bibitem{Mueller2017}
Mueller, J. P. B., Rubin, N. A., Devlin, R. C., Groever, B. \& Capasso, F. Metasurface polarization optics: independent phase control of arbitrary orthogonal states of polarization. \emph{Phys. Rev. Lett.} \textbf{118}, 113901 (2017).

\bibitem{Arbabi2015}
Arbabi, A., Horie, Y., Bagheri, M. \& Faraon, A. Dielectric metasurfaces for complete control of phase and polarization with subwavelength spatial resolution and high transmission. \emph{Nat. Nanotechnol.} \textbf{10}, 937--943 (2015).

\bibitem{Rubin2021}
Rubin, N. A., Zaidi, A., Dorrah, A. H., Shi, Z. \& Capasso, F. Jones matrix holography with metasurfaces. \emph{Sci. Adv.} \textbf{7}, eabg7488 (2021).

\bibitem{Dorrah2021NP}
Dorrah, A. H., Rubin, N. A., Zaidi, A., Tamagnone, M. \& Capasso, F. Metasurface optics for on-demand polarization transformations along the optical path. \emph{Nat. Photon.} \textbf{15}, 287--296 (2021).

\bibitem{Dorrah2021NC}
Dorrah, A. H., Rubin, N. A., Tamagnone, M., Zaidi, A. \& Capasso, F. Structuring total angular momentum of light along the propagation direction with polarization-controlled meta-optics. \emph{Nat. Commun.} \textbf{12}, 6249 (2021).

\bibitem{Liu2024PRJ}
Liu, C. et al. Multi-plane vectorial holography based on a height tunable metasurface fabricated by femtosecond laser direct writing. \emph{Photon. Res.} \textbf{12}, 2158--2165 (2024).

\bibitem{Wang2024NC}
Wang, D. et al. Decimeter-depth and polarization addressable color 3D meta-holography. \emph{Nat. Commun.} \textbf{15}, 8242 (2024).

\bibitem{Tan2026LSA}
Tan, L. et al. Longitudinally engineered metasurfaces for 3D vectorial holography. \emph{Light Sci. Appl.} \textbf{15}, 36 (2026).

\bibitem{Zong2026LPR}
Zong, S. et al. Three-dimensional vectorial holography with full polarization control via end-to-end on-chip metasurface. \emph{Laser Photon. Rev.} e71812, https://doi.org/10.1002/lpor.71812 (2026).

\bibitem{Intaravanne2022}
Intaravanne, Y. et al. Color-selective three-dimensional polarization structures. \emph{Light Sci. Appl.} \textbf{11}, 302 (2022).

\bibitem{Li2024ACS}
Li, Y. et al. Multiple multicolored 3D polarization knots arranged along light propagation. \emph{ACS Photonics} \textbf{11}, 4380--4389 (2024).

\bibitem{MataCervera2025}
Mata-Cervera, N. et al. Tailoring propagation-invariant topology of optical skyrmions with dielectric metasurfaces. \emph{Nanophotonics} \textbf{14}, 4069--4077 (2025).

\bibitem{Fu2026}
Fu, Y. et al. Free-form flexible metasurfaces robustly generating microwave skyrmions. \emph{Adv. Mater.} \textbf{38}, e18268 (2026).

\bibitem{Park2026}
Park, D., Song, A., Chung, H. \& Kim, S. Inverse-designed metasurfaces for compact optical skyrmion generation with high topological fidelity. Preprint at https://arxiv.org/abs/2603.23922 (2026).

\bibitem{Xie2026}
Xie, R., Mata-Cervera, N., Xie, X. \& Shen, Y. Metasurfaces and metadevices for topological electromagnetic waves. \emph{Adv. Phys. Res.} \textbf{5}, e00209 (2026).

\bibitem{Allen1992}
Allen, L., Beijersbergen, M. W., Spreeuw, R. J. C. \& Woerdman, J. P. Orbital angular momentum of light and the transformation of Laguerre--Gaussian laser modes. \emph{Phys. Rev. A} \textbf{45}, 8185--8189 (1992).

\bibitem{Siegman1986}
Siegman, A. E. \emph{Lasers} (University Science Books, 1986).

\bibitem{Skyrme1961}
Skyrme, T. H. R. A non-linear field theory. \emph{Proc. R. Soc. Lond. A} \textbf{260}, 127--138 (1961).

\bibitem{Milnor1965}
Milnor, J. W. \emph{Topology from the Differentiable Viewpoint} (University Press of Virginia, 1965).

\bibitem{Hatcher2002}
Hatcher, A. \emph{Algebraic Topology} (Cambridge Univ. Press, 2002).

\bibitem{Nadarloo2026}
Nadarloo, A., Hajiahmadi, M. J. \& Soleimani, M. A single passive terahertz metasurface implementing eight Boolean logic gates through polarization multiplexing. \emph{Sci. Rep.} https://doi.org/10.1038/s41598-026-69471-w (2026).

\bibitem{Wang2024LSA}
Wang, A. A. et al. Topological protection of optical skyrmions through complex media. \emph{Light Sci. Appl.} \textbf{13}, 314 (2024).

\bibitem{Goodman2005}
Goodman, J. W. \emph{Introduction to Fourier Optics} 3rd edn (Roberts \& Company, 2005).

\bibitem{Freudenthal1942}
Freudenthal, H. Simplizialzerlegungen von beschr{\"a}nkter Flachheit. \emph{Ann. Math.} \textbf{43}, 580--582 (1942).

\end{thebibliography}
\end{document}


\maketitle
\tableofcontents
\clearpage

\section{Closed-form hopfion beams and their topological order}
\label{sn:theory}
This note derives the results stated in the main-text sections on the target field, the hopfion order and the two phase maps. It uses only the paraxial wave equation, two properties of the Laguerre polynomials and the elementary degree theory of maps~\cite{Milnor1965,Hatcher2002}.

\subsection*{The result in brief}
The focal fields $A_f=\rho^{\ell}e^{\mathrm{i}\ell\theta}e^{-\rho^2}$ and $B_f=\sum_pc_pL_p(2\rho^2)e^{-\rho^2}$ propagate in closed form (Part~1). In co-moving coordinates all of space becomes a solid cylinder whose boundary is the far field and the region far from the axis. Under conditions (i)--(iii) the boundary is sent to a disc of states around the $x$-polarized pole and a circle, a set of zero volume that does not divide the three-sphere. The signed count of preimages is then the same for almost every value, and it is the Skyrme number. Counting the preimages of one $y$-polarized value gives $\ell$ per ring, and the argument principle gives $-n$ for the sum of the ring indices, so $|N_S|=n\ell$ (Part~2). The states that the far field never reaches have closed preimages, and these link exactly $n\ell$ times (corollary after Step~12). Part~3 turns the fields into the two masks.

\subsection*{Symbols and notation}
The symbols below are used in this note, in the main text and in the other Supplementary Notes. A few letters have two meanings in different parts of the work; both meanings are listed, and the context always makes clear which one is meant.

\medskip
\noindent\textbf{Coordinates and beam parameters}\par\smallskip
{\small\noindent
\begin{tabular}{@{}>{\raggedright\arraybackslash}p{0.24\linewidth}>{\raggedright\arraybackslash}p{0.70\linewidth}@{}}
\toprule
Symbol & Meaning\\
\midrule
$x,\ y,\ z$ & Cartesian coordinates. The beam travels along $z$; $z$ is measured from the exit face of the metasurface, which lies in the plane $z=0$.\\
$r,\ \theta$ & Polar coordinates in a transverse plane: $x=r\cos\theta$, $y=r\sin\theta$. $r$ is the distance from the beam axis and $\theta$ the azimuthal angle.\\
$\mathrm{i}$ & Imaginary unit, $\mathrm{i}^2=-1$.\\
$t$ & Complex variable of the polynomial $Q(t)$ (Steps~T5 and 12--15).\\
$\omega,\ \lambda,\ k$ & Angular frequency, free-space wavelength and wavenumber, $k=2\pi/\lambda$.\\
$z_0$ & Distance from the metasurface to the focal plane, where the target field is prescribed.\\
$w_0$ & Waist of the target: the beam radius in the focal plane.\\
$z_R$ & Rayleigh range, $z_R=kw_0^2/2=\pi w_0^2/\lambda$: the distance over which the beam stays nearly as narrow as at its waist.\\
$\rho$ & Distance from the axis in units of the waist, $\rho=r/w_0$.\\
$\tau$ & Distance from the focal plane in units of the Rayleigh range, $\tau=(z-z_0)/z_R$.\\
$\nu,\ \nu^{*}$ & Complex beam parameter $\nu=1+\mathrm{i}\tau$ and its complex conjugate $\nu^{*}=1-\mathrm{i}\tau$.\\
$w(z)$ & Local beam radius, $w(z)=w_0|\nu|=w_0(1+\tau^2)^{1/2}$.\\
$\psi$ & Gouy angle, $\psi=\arg\nu=\arctan\tau$. It runs from $-\pi/2$ (far before the focus) to $+\pi/2$ (far behind it).\\
$\xi$ & Co-moving distance from the axis, $\xi=r/w(z)=\rho/|\nu|$: the distance in units of the local beam radius.\\
$\zeta$ & Distance from the sheet to the focus in units of the Rayleigh range, $\zeta=z_0/z_R$. On the sheet, $\tau=-\zeta$.\\
$R$ & Focal length of the lens term common to both channels, $R=z_0(1+\zeta^{-2})$.\\
$w_{\rm in}$ & Waist of the illuminating Gaussian beam at the sheet.\\
\bottomrule
\end{tabular}}

\medskip
\noindent\textbf{Fields and polarization}\par\smallskip
{\small\noindent
\begin{tabular}{@{}>{\raggedright\arraybackslash}p{0.24\linewidth}>{\raggedright\arraybackslash}p{0.70\linewidth}@{}}
\toprule
Symbol & Meaning\\
\midrule
$E_x,\ E_y,\ E_z$ & Complex amplitudes of the three components of the electric field.\\
$B\equiv E_x,\ A\equiv E_y$ & The two transverse components from which the polarization texture is built.\\
$A_f,\ B_f$ & The two target components in the focal plane $z=z_0$.\\
$u$ & Slowly varying envelope of one field component, which is written as $u\,e^{\mathrm{i} kz}$.\\
$u_{\rm in}$ & Illuminating field at the sheet.\\
$g$ & Fundamental Gaussian beam, $g=\nu^{-1}e^{-\rho^2/\nu}$; it never vanishes.\\
$A/g,\ B/g$ & Reduced components: the fields divided by the common Gaussian factor $g$.\\
$\phi_x,\ \phi_y$ & Transmission phases that a metasurface cell imposes on $x$- and $y$-polarized light.\\
$J$ & Jones matrix of a cell.\\
$S_0$ & Total transverse intensity, $S_0=|A|^2+|B|^2$.\\
$\mathbf n=(n_x,n_y,n_z)$ & Normalized Stokes vector: the local polarization as a point on the Poincar\'e sphere. $n_z=+1$ is $y$-polarized light ($B=0$) and $n_z=-1$ is $x$-polarized light ($A=0$).\\
$w$ & Complex ratio that labels a polarization state, $w=E_y/E_x$; its preimage is the set of points where $A-wB=0$. In the main text the traced states are labeled by $w=\tilde A/\tilde B$.\\
$\tilde A,\ \tilde B,\ \kappa_{AB}$ & Power-balanced pair, $\tilde A=A$ and $\tilde B=\kappa_{AB}B$, with $\kappa_{AB}$ the ratio of the root-mean-square amplitudes of $A$ and $B$.\\
$\mathbf u$ (bold) & Normalized four-component field, a point on the three-sphere $S^3$ (Step~T8).\\
$S^2,\ S^3$ & Unit sphere in three dimensions (the Poincar\'e sphere) and in four dimensions.\\
\bottomrule
\end{tabular}}

\medskip
\noindent\textbf{Design parameters and topology}\par\smallskip
{\small\noindent
\begin{tabular}{@{}>{\raggedright\arraybackslash}p{0.24\linewidth}>{\raggedright\arraybackslash}p{0.70\linewidth}@{}}
\toprule
Symbol & Meaning\\
\midrule
$\ell$ & Vortex order, a positive integer: the number of times the phase of $A$ turns around the beam axis.\\
$n$ & Radial order of the ring component, a positive integer: the highest $p$ with $c_p\neq0$.\\
$p$ & Radial mode index, $0\le p\le n$. In the main text, $p$ also denotes the lattice pitch.\\
$c_p$ & Real coefficients of the ring component, with $c_n\neq0$.\\
$L_p$ & Laguerre polynomial of degree $p$ (Step~T7).\\
$\mathrm{LG}_p^{l}$ & Laguerre--Gauss beam with radial index $p$ and azimuthal index $l$.\\
$\kappa,\ r_c$ & Parameter and radius of the single-ring design, $\kappa=w_0^2/r_c^2$.\\
$F(x)$ & Far-field profile of the ring component, with $x=\xi^2$ (Step~8). In the main text, $F$ also denotes the field overlap; bold $\mathbf F$ is the field of the Whitehead integral (Supplementary Note~\ref{sn:first}).\\
$Q(t)$ & On-axis polynomial $Q(t)=\sum_pc_pt^p$ of a complex variable $t$; on the axis, $t=e^{-2\mathrm{i}\psi}$.\\
$m_0$ & Number of roots of $Q$ with $|t|<1$, counted with multiplicity.\\
$N_S$ & Skyrme number, equation~(7) of the main text.\\
$\deg(y),\ \deg$ & Signed number of points of space that the map $\mathbf u$ sends to the point $y$ of $S^3$, and its common value (the degree of the map).\\
$J_{\mathbf u}$ & $3\times3$ Jacobian matrix of $\mathbf u$ in oriented local coordinates (Step~10).\\
$Lk$ & Gauss linking number of two closed curves, equation~(3) of the main text.\\
$\varepsilon_j$ & Index ($+1$ or $-1$) of the $j$-th zero of $B/g$ (Step~T3). Without a subscript, $\varepsilon$ is the level of the regularizing background.\\
$\sigma_0$ & A fixed sign, $+1$ or $-1$, set by the orientation of the coordinates.\\
$H$ & Open half-plane $\{\xi>0,\ |\psi|<\pi/2\}$ of the co-moving coordinates.\\
$\Delta,\ C,\ K$ & Disc, circle and their union: the set of values that $\mathbf u$ takes on the boundary of space (Step~9).\\
$R(\xi),\ R_{\max}$ & Far-field ratio $|A/B|$ and its largest value.\\
$X$ & A large value of $\xi$ used to close the counting rectangle (Step~12).\\
$D$ & Derivative operator $D=\partial_x+\mathrm{i}\partial_y$ (Step~5).\\
$a,\ F_a,\ \nu_a,\ \beta$ & Width parameter of a general Gaussian and the Gaussian beam built from it (Step~3). Bold $\mathbf a$ is the vector potential of the Whitehead integral.\\
$s,\ q,\ a_s$ & Auxiliary variables of the Laguerre generating function (Step~6).\\
\bottomrule
\end{tabular}}

\medskip
\noindent\textbf{Numerical measures and device (other Supplementary Notes)}\par\smallskip
{\small\noindent
\begin{tabular}{@{}>{\raggedright\arraybackslash}p{0.24\linewidth}>{\raggedright\arraybackslash}p{0.70\linewidth}@{}}
\toprule
Symbol & Meaning\\
\midrule
$Q_W,\ \mathbf F,\ \mathbf a$ & Whitehead integral, its field $\mathbf F$ and a vector potential $\mathbf a$ with $\nabla\times\mathbf a=\mathbf F$.\\
$\varepsilon$ & Level of the incoherent $x$-polarized background, relative to the peak intensity.\\
$\mathbf n_\varepsilon$ & Texture after the background is added.\\
$h$ & Grid step of the numerical integration; for the meta-atom, the quartz spacer thickness.\\
$F,\ F(z)$ & Field overlap between two fields over a volume, and over a single transverse plane.\\
$L_x,\ L_y,\ L_w,\ r$ & Arm lengths, arm width and disk radius of the meta-atom.\\
\bottomrule
\end{tabular}}

\subsection*{How this note is organized}
The derivation answers three questions, in this order.
\begin{enumerate}
\item How do the two target components travel through space? (Steps~1--7.) The answer is an exact formula for both fields at every point, equation~\eqref{eq:s_lift}.
\item Why is the hopfion order exactly $n\ell$? (Steps~8--13.) The answer counts, with signs, how often the field reaches one particular polarization state, shows that this count equals the Skyrme number, and then shows that it is also the linking number of closed polarization loops.
\item How do the formulas become two metasurface masks? (Steps~14--17.) Worked examples with one, two and three rings and the design procedure are given.
\end{enumerate}
Before these three parts, Steps~T1--T8, preceded by a remark on the Gouy phase, collect the few tools that are needed. A reader familiar with polarization optics, complex numbers, winding numbers and the degree of a map can go directly to Step~1.

\paragraph{A remark on the Gouy phase.} A focused beam does not accumulate exactly the phase $kz$ of a plane wave: in passing through the focus it acquires an extra phase, the Gouy phase. The fundamental Gaussian beam carries the Gouy factor $e^{-\mathrm{i}\psi}$, where $\psi=\arctan[(z-z_0)/z_R]$ changes by $\pi$ from far before the focus to far behind it; a mode of total order $N$ carries the factor $e^{-\mathrm{i}(N+1)\psi}$ (Steps~5 and 6). The different Gouy phases of the modes that make up $B$ change their relative weights along the beam, and this is what shapes the dark rings in three dimensions (Step~6).

\subsection*{Mathematical tools}

\paragraph{Step T1. Complex numbers in polar form.} Every complex number can be written as $z=|z|\,e^{\mathrm{i}\alpha}$, where $|z|$ is its modulus (its length) and $\alpha=\arg z$ its argument (its direction), with $e^{\mathrm{i}\alpha}=\cos\alpha+\mathrm{i}\sin\alpha$. The argument is defined only up to multiples of $2\pi$. When two complex numbers are multiplied, the moduli multiply and the arguments add; when they are divided, the moduli divide and the arguments subtract. The complex conjugate $z^{*}=|z|e^{-\mathrm{i}\alpha}$ has the same modulus and the opposite argument, and $zz^{*}=|z|^2$. A number $e^{\mathrm{i}\alpha}$ always has modulus one. Two special cases are used repeatedly: $e^{\mathrm{i}\pi}=e^{-\mathrm{i}\pi}=-1$ and $e^{\pm\mathrm{i}\pi/2}=\pm\mathrm{i}$.

\paragraph{Step T2. Counting turns.} Let a complex quantity $f$ vary continuously along a path and never vanish on it. Its argument can then be followed continuously, without jumps of $2\pi$, from the start of the path to its end; the difference between the final and the initial value is the \emph{net change of the argument}. If the path is closed and $f$ returns to its initial value, this net change is an integer multiple of $2\pi$. The integer is the \emph{number of turns} of $f$ along the path: positive if $f$ turns counterclockwise in the complex plane, negative if it turns clockwise. For example, $f=e^{\mathrm{i} m\theta}$ with an integer $m$ makes $m$ turns as $\theta$ runs from 0 to $2\pi$. The net change along a path made of several pieces is the sum of the net changes along the pieces.

\paragraph{Step T3. Zeros of a complex function of two variables and their index.} Let $f(\xi,\psi)$ be a smooth complex function of two real variables, and let $f$ vanish at a point $P$ and nowhere else nearby. The \emph{index} of the zero is the number of turns of $f$ along a small circle around $P$, traversed counterclockwise in the $(\xi,\psi)$ plane. Near a \emph{simple} zero, $f$ is well approximated by its linear part,
\begin{equation}
f\approx a\,\delta\xi+b\,\delta\psi ,
\end{equation}
where $\delta\xi$ and $\delta\psi$ are the displacements from $P$ and $a=\partial_\xi f$, $b=\partial_\psi f$ are complex numbers. In terms of real and imaginary parts, this linear map sends $(\delta\xi,\delta\psi)$ to $(\mathrm{Re}f,\mathrm{Im}f)$ with the matrix
\begin{equation}
\mathcal J=\begin{pmatrix}\mathrm{Re}\,a & \mathrm{Re}\,b\\ \mathrm{Im}\,a & \mathrm{Im}\,b\end{pmatrix},\qquad \det\mathcal J=\mathrm{Im}(a^{*}b).
\end{equation}
A zero is simple when $\det\mathcal J\neq0$. The small circle around $P$ is then mapped onto an ellipse around the origin, which is traversed counterclockwise if $\det\mathcal J>0$ and clockwise if $\det\mathcal J<0$. The index of a simple zero is therefore $+1$ or $-1$, equal to the sign of $\det\mathcal J$. For example, $f=\delta\xi+\mathrm{i}\,\delta\psi$ has index $+1$, and $f=\delta\xi-\mathrm{i}\,\delta\psi$ has index $-1$. A zero that is not simple can be split by an arbitrarily small change of $f$ into simple zeros whose indices add up to its own index.

\paragraph{Step T4. Counting zeros inside a closed curve.} Let $f$ be nonzero on a closed curve that bounds a region of the $(\xi,\psi)$ plane, and let $f$ have finitely many zeros inside it. Then
\begin{equation}
\text{number of turns of $f$ along the curve (counterclockwise)}=\text{sum of the indices of the zeros inside}.
\label{eq:s_count}
\end{equation}
To see this, divide the region into many small cells, each traversed counterclockwise. Every inner edge belongs to two neighboring cells and is traversed once in each direction, so its contributions cancel, and the sum over all cells equals the number of turns along the outer curve. A small cell without a zero contributes nothing, because $f$ is nearly constant in it and cannot turn around the origin. A small cell that contains one zero contributes the index of that zero, by Step~T3.

\paragraph{Step T5. A polynomial on the unit circle.} Let $Q(t)=c_0+c_1t+\dots+c_nt^n$ with $c_n\neq0$. By the fundamental theorem of algebra it factorizes as $Q(t)=c_n(t-t_1)(t-t_2)\cdots(t-t_n)$, where $t_1,\dots,t_n$ are its roots. By Step~T1, $\arg Q=\arg c_n+\sum_k\arg(t-t_k)$. Now let $t=e^{\mathrm{i}\alpha}$ run once counterclockwise around the unit circle and suppose that no root lies on the circle. For a root inside the circle ($|t_k|<1$), the vector $t-t_k$ points from $t_k$ to a point that goes once around $t_k$, so $\arg(t-t_k)$ increases by $2\pi$. For a root outside the circle ($|t_k|>1$), the point $t$ never goes around $t_k$, and $\arg(t-t_k)$ returns to its initial value. Therefore
\begin{equation}
\text{number of turns of $Q$ along the unit circle}=m_0=\text{number of roots of $Q$ with $|t|<1$},
\label{eq:s_roots}
\end{equation}
where a repeated root is counted as often as it occurs in the factorization (with multiplicity). This is the argument principle for polynomials.

\paragraph{Step T6. A dominant term fixes the turning.} Let $f=f_0+f_1$ along a path, where $|f_1|<|f_0|$ everywhere on it. Then $f/f_0=1+f_1/f_0$ lies inside the disc of radius one around the point 1 of the complex plane. This disc lies entirely in the half-plane $\mathrm{Re}>0$, so the argument of $f/f_0$ stays between $-\pi/2$ and $\pi/2$ and can never make a turn. Since $\arg f=\arg f_0+\arg(f/f_0)$, the net change of $\arg f$ equals the net change of $\arg f_0$ plus the net change of $\arg(f/f_0)$, and the latter is zero whenever $f/f_0$ is real and positive at both ends of the path. In short, a term that is larger than all the others decides how $f$ turns.

\paragraph{Step T7. Laguerre polynomials.} The Laguerre polynomials are
\begin{equation}
\begin{aligned}
L_p(y)&=\sum_{m=0}^{p}\binom{p}{m}\frac{(-y)^m}{m!},\\
L_0&=1,\quad L_1(y)=1-y,\quad L_2(y)=1-2y+\tfrac12y^2,\quad L_3(y)=1-3y+\tfrac32y^2-\tfrac16y^3 .
\end{aligned}
\end{equation}
Two properties are used below. First, $L_p(0)=1$ for every $p$, and the highest power of $L_p(y)$ is $(-1)^py^p/p!$. Second, all of them are collected in the generating function~\cite{DLMF}
\begin{equation}
\sum_{p=0}^{\infty}L_p(y)\,s^p=\frac{1}{1-s}\exp\Bigl(-\frac{ys}{1-s}\Bigr),\qquad |s|<1 ,
\label{eq:s_genfun}
\end{equation}
which holds for real and complex $s$. It can be checked term by term: with $1/(1-s)=1+s+s^2+\dots$ and $-ys/(1-s)=-ys-ys^2-\dots$, the right-hand side is $1+(1-y)s+(1-2y+\tfrac12y^2)s^2+\dots$, which reproduces $L_0$, $L_1$ and $L_2$.

\paragraph{Step T8. The three-sphere and the degree of a map.} A pair of complex numbers $(A,B)$ with $|A|^2+|B|^2=1$ is a point of the three-sphere $S^3$: the set of points $(a_1,a_2,a_3,a_4)$, with $a_1=\mathrm{Re}A$, $a_2=\mathrm{Im}A$, $a_3=\mathrm{Re}B$ and $a_4=\mathrm{Im}B$, at unit distance from the origin of four-dimensional space. $S^3$ is a closed three-dimensional surface of total volume $2\pi^2$. Such a pair carries three real numbers: the polarization state, fixed by the ratio $A/B$ (two numbers, a point on the Poincar\'e sphere), and the common phase of $A$ and $B$ (one number).

Let a map $\mathbf u$ send each point of a three-dimensional region to a point of $S^3$. For a point $y$ of $S^3$, the \emph{signed count} $\deg(y)$ is the number of points of the region that are sent to $y$, each counted $+1$ if the map preserves orientation there and $-1$ if it reverses it (the sign of its $3\times3$ Jacobian determinant in oriented local coordinates on the region and on $S^3$). The Skyrme number is the total signed volume of $S^3$ covered by $\mathbf u$, divided by $2\pi^2$. Since every point $y$ is covered $\deg(y)$ times, the Skyrme number is the average of $\deg(y)$ over $S^3$. If $\deg(y)$ has the same value for all $y$ except a set of zero volume, the Skyrme number equals this value and is an integer, called the \emph{degree} of the map.

The count $\deg(y)$ can change only in one way. As $y$ moves, its preimages move continuously. Inside the region, preimages can appear or disappear only in pairs of opposite sign, like two roots of a quadratic equation that merge and leave the real axis; this leaves the signed count unchanged. The count can change only if a preimage leaves through the boundary of the region, that is, if $y$ crosses the set of values that $\mathbf u$ takes on the boundary~\cite{Milnor1965}. The whole derivation below is built on this fact.

\subsection*{Part 1: how the two components propagate}

\paragraph{Step 1. The paraxial wave equation.} Each transverse field component $E$ obeys the Helmholtz equation $\partial_x^2E+\partial_y^2E+\partial_z^2E+k^2E=0$. For a beam that travels along $z$ we write $E=u(x,y,z)\,e^{\mathrm{i} kz}$, where the envelope $u$ changes slowly along $z$ compared with the carrier $e^{\mathrm{i} kz}$. The product rule gives $\partial_z^2E=(\partial_z^2u+2\mathrm{i} k\,\partial_zu-k^2u)\,e^{\mathrm{i} kz}$, and the Helmholtz equation becomes $\partial_x^2u+\partial_y^2u+\partial_z^2u+2\mathrm{i} k\,\partial_zu=0$. For a slowly varying envelope, $\partial_z^2u$ is much smaller than $2k\,\partial_zu$ and is neglected. This leaves the paraxial wave equation~\cite{Siegman1986}
\begin{equation}
2\mathrm{i} k\,\partial_z u+\partial_x^2u+\partial_y^2u=0 .
\label{eq:s_parax}
\end{equation}
Everything below is exact within equation~\eqref{eq:s_parax}; no further approximation is made.

\paragraph{Step 2. Three properties of the paraxial equation.} (a) The equation is linear: a sum of solutions, or a solution multiplied by a constant, is again a solution. (b) Its coefficients do not depend on $x$ or $y$. Therefore, if $u$ is a solution, so are $\partial_xu$ and $\partial_yu$, because differentiating the whole equation with respect to $x$ or $y$ gives the same equation for the derivative. (c) The equation contains only the first derivative with respect to $z$. A solution is therefore fixed completely by its values in one plane $z=z_0$: once the field in the focal plane is prescribed, the field everywhere else is determined. To propagate a prescribed focal field, it is therefore enough to find any solution that equals it at $z=z_0$.

\paragraph{Step 3. A Gaussian beam of any width.} Let $a>0$ be a real constant, and let
\begin{equation}
F_a=\frac{1}{\nu_a}\exp\Bigl(-\frac{ar^2}{\nu_a}\Bigr),\qquad \nu_a=1+\frac{2\mathrm{i} a(z-z_0)}{k},\qquad r^2=x^2+y^2 .
\label{eq:s_gauss}
\end{equation}
At $z=z_0$, $\nu_a=1$ and $F_a=e^{-ar^2}$, a Gaussian of width $a^{-1/2}$. We check that $F_a$ solves equation~\eqref{eq:s_parax}. Write $\beta=a/\nu_a$, so that $F_a=\nu_a^{-1}e^{-\beta r^2}$. Since $\partial_z\nu_a=2\mathrm{i} a/k$,
\begin{equation}
\partial_z\Bigl(\frac1{\nu_a}\Bigr)=-\frac{2\mathrm{i}}{k}\,\frac{\beta}{\nu_a},\qquad \partial_z\beta=-\frac{2\mathrm{i}}{k}\,\beta^2,\qquad\text{so}\qquad \partial_zF_a=\frac{2\mathrm{i}}{k}\bigl(\beta^2r^2-\beta\bigr)F_a .
\end{equation}
Multiplying by $2\mathrm{i} k$ gives $2\mathrm{i} k\,\partial_zF_a=(4\beta-4\beta^2r^2)F_a$. For the transverse part, $\partial_xe^{-\beta r^2}=-2\beta x\,e^{-\beta r^2}$ and $\partial_x^2e^{-\beta r^2}=(4\beta^2x^2-2\beta)\,e^{-\beta r^2}$, and in the same way for $y$; hence $(\partial_x^2+\partial_y^2)F_a=(4\beta^2r^2-4\beta)F_a$. The two parts cancel exactly, so $F_a$ solves the paraxial equation.

\paragraph{Step 4. The fundamental beam and the co-moving coordinates.} For $a=w_0^{-2}$, $F_a$ is the fundamental Gaussian beam of waist $w_0$ at $z=z_0$. With $\rho=r/w_0$, the Rayleigh range $z_R=kw_0^2/2$ and $\tau=(z-z_0)/z_R$, one has $\nu_a=1+\mathrm{i}\tau\equiv\nu$ and
\begin{equation}
g=\frac{1}{\nu}\,e^{-\rho^2/\nu}.
\label{eq:s_g}
\end{equation}
Write $\nu$ in polar form (Step~T1), $\nu=|\nu|e^{\mathrm{i}\psi}$, with
\begin{equation}
|\nu|=(1+\tau^2)^{1/2},\qquad \psi=\arctan\tau .
\end{equation}
Since $1/\nu=\nu^{*}/|\nu|^2=(1-\mathrm{i}\tau)/(1+\tau^2)$, the modulus of $g$ is $|g|=|\nu|^{-1}\exp[-\rho^2/(1+\tau^2)]=|\nu|^{-1}\exp[-r^2/w(z)^2]$ with $w(z)=w_0|\nu|$. The beam therefore keeps its Gaussian shape while its radius grows as $w(z)$, which is why $|\nu|$ is called the beam expansion. The angle $\psi$ is the Gouy angle; as $z$ runs from $-\infty$ to $+\infty$, $\tau$ does the same and $\psi$ increases from $-\pi/2$ to $\pi/2$. We measure the distance from the axis in units of the local beam radius,
\begin{equation}
\xi=\frac{r}{w(z)}=\frac{\rho}{|\nu|}.
\end{equation}
Four identities, each of which follows directly from $\nu=1+\mathrm{i}\tau=|\nu|e^{\mathrm{i}\psi}$, are used below:
\begin{equation}
\text{(I1)}\ \ \frac{\rho}{\nu}=\xi\,e^{-\mathrm{i}\psi},\qquad \text{(I2)}\ \ \frac{\nu^{*}}{\nu}=e^{-2\mathrm{i}\psi},\qquad \text{(I3)}\ \ \nu\nu^{*}=|\nu|^2=1+\tau^2,\qquad \text{(I4)}\ \ 2-\nu^{*}=\nu .
\end{equation}

\paragraph{Step 5. The vortex component $A$.} We need the solution whose focal field is $A_f=\rho^{\ell}e^{\mathrm{i}\ell\theta}e^{-\rho^2}$. Since $x+\mathrm{i} y=re^{\mathrm{i}\theta}$, this focal field equals $w_0^{-\ell}(x+\mathrm{i} y)^{\ell}e^{-r^2/w_0^2}$. The factor $(x+\mathrm{i} y)^{\ell}$ is produced by the derivative operator
\begin{equation}
D=\partial_x+\mathrm{i}\,\partial_y .
\end{equation}
Two facts follow from the chain rule. First, $D(x+\mathrm{i} y)=1+\mathrm{i}\cdot\mathrm{i}=0$, and therefore also $D(x+\mathrm{i} y)^m=0$. Second, for any function $h$ of $r^2$, $D\,h(r^2)=h'(r^2)\,(2x+2\mathrm{i} y)=2(x+\mathrm{i} y)\,h'(r^2)$. By the product rule, $D[(x+\mathrm{i} y)^mh(r^2)]=2(x+\mathrm{i} y)^{m+1}h'(r^2)$. For $h=e^{-cr^2}$, with any constant $c$, $h'=-c\,e^{-cr^2}$, so each application of $D$ multiplies by $-2c(x+\mathrm{i} y)$:
\begin{equation}
D^{\ell}e^{-cr^2}=(-2c)^{\ell}(x+\mathrm{i} y)^{\ell}e^{-cr^2}.
\end{equation}
Applying this to the Gaussian of Step~3, with $c=a/\nu_a$ (which does not depend on $x$ or $y$), gives
\begin{equation}
D^{\ell}F_a=\frac{1}{\nu_a}\Bigl(-\frac{2a}{\nu_a}\Bigr)^{\ell}(x+\mathrm{i} y)^{\ell}e^{-ar^2/\nu_a}.
\end{equation}
By Step~2(b), $D^{\ell}F_a$ is a solution, and by Step~2(a) so is $D^{\ell}F_a/(-2a)^{\ell}=(x+\mathrm{i} y)^{\ell}\nu_a^{-(\ell+1)}e^{-ar^2/\nu_a}$, whose value at $z=z_0$ is $(x+\mathrm{i} y)^{\ell}e^{-ar^2}$. With $a=w_0^{-2}$ and the factor $w_0^{-\ell}$, the solution with focal field $A_f$ is
\begin{equation}
A=\frac{(x+\mathrm{i} y)^{\ell}}{w_0^{\ell}}\,\frac{1}{\nu^{\ell+1}}\,e^{-\rho^2/\nu}=g\,\Bigl(\frac{\rho}{\nu}\Bigr)^{\ell}e^{\mathrm{i}\ell\theta}.
\end{equation}
With identity (I1),
\begin{equation}
\frac{A}{g}=\xi^{\ell}e^{\mathrm{i}\ell\theta}e^{-\mathrm{i}\ell\psi}.
\label{eq:s_A}
\end{equation}
In words: the vortex keeps its shape in the co-moving coordinate $\xi$, vanishes only on the axis, turns $\ell$ times around it, and acquires the extra Gouy phase $e^{-\mathrm{i}\ell\psi}$.

\paragraph{Step 6. The ring component $B$.} We need the solution whose focal field is the ring component $B_f$ of equation~(4) of the main text, a sum of the radial modes $L_p(2\rho^2)e^{-\rho^2}$ with the weights $c_p$. By linearity, it is enough to propagate each radial mode $L_p(2\rho^2)e^{-\rho^2}$ separately. The generating function~\eqref{eq:s_genfun} propagates all of them at once. Let $s$ be a real number between $-1$ and $1$. Setting $y=2\rho^2$ in equation~\eqref{eq:s_genfun} and multiplying by $e^{-\rho^2}$ gives
\begin{equation}
\sum_{p=0}^{\infty}s^p\,L_p(2\rho^2)\,e^{-\rho^2}=\frac{1}{1-s}\exp\Bigl[-\rho^2\Bigl(1+\frac{2s}{1-s}\Bigr)\Bigr]=\frac{1}{1-s}\,e^{-a_sr^2},\qquad a_s=\frac{1+s}{(1-s)\,w_0^2}.
\end{equation}
The right-hand side is a Gaussian with the positive width parameter $a_s$, multiplied by a constant. By Step~3 and Step~2(a) it propagates as $(1-s)^{-1}F_{a_s}$. We now simplify this expression. First, since $2a_s(z-z_0)/k=\tau(1+s)/(1-s)$,
\begin{equation}
(1-s)\,\nu_{a_s}=(1-s)+\mathrm{i}\tau(1+s)=(1+\mathrm{i}\tau)-s(1-\mathrm{i}\tau)=\nu-s\nu^{*}.
\end{equation}
Define
\begin{equation}
q=s\,\frac{\nu^{*}}{\nu}=s\,e^{-2\mathrm{i}\psi}
\end{equation}
by identity (I2); note that $|q|=|s|<1$. Then the prefactor is
\begin{equation}
\frac{1}{(1-s)\,\nu_{a_s}}=\frac{1}{\nu-s\nu^{*}}=\frac{1}{\nu}\,\frac{1}{1-q}.
\end{equation}
The exponent is $a_sr^2/\nu_{a_s}=\rho^2(1+s)/[(1-s)\nu_{a_s}]=\rho^2(1+s)/(\nu-s\nu^{*})$. We show that it equals $\rho^2/\nu+2\xi^2q/(1-q)$. Using $\xi^2=\rho^2/(\nu\nu^{*})$ from (I3) and the definition of $q$,
\begin{equation}
\frac{2\xi^2q}{1-q}=\frac{2\rho^2}{\nu\nu^{*}}\,\frac{s\nu^{*}/\nu}{(\nu-s\nu^{*})/\nu}=\frac{2\rho^2s}{\nu\,(\nu-s\nu^{*})},
\end{equation}
so that
\begin{equation}
\frac{\rho^2}{\nu}+\frac{2\xi^2q}{1-q}=\frac{\rho^2}{\nu}\,\frac{\nu-s\nu^{*}+2s}{\nu-s\nu^{*}}=\frac{\rho^2}{\nu}\,\frac{\nu(1+s)}{\nu-s\nu^{*}}=\frac{\rho^2(1+s)}{\nu-s\nu^{*}},
\end{equation}
where we used $\nu-s\nu^{*}+2s=\nu+s(2-\nu^{*})=\nu+s\nu$ by (I4). Collecting the prefactor and the exponent, and using equation~\eqref{eq:s_g}, the propagated field is
\begin{equation}
\frac{1}{\nu}\,\frac{1}{1-q}\,\exp\Bigl(-\frac{\rho^2}{\nu}-\frac{2\xi^2q}{1-q}\Bigr)=g\cdot\frac{1}{1-q}\exp\Bigl(-\frac{2\xi^2q}{1-q}\Bigr)=g\sum_{p=0}^{\infty}q^p\,L_p(2\xi^2),
\end{equation}
where the last step is the generating function~\eqref{eq:s_genfun} again, now with $y=2\xi^2$ and the complex variable $q$. Since $q^p=s^pe^{-2\mathrm{i} p\psi}$, we have found
\begin{equation}
\sum_{p=0}^{\infty}s^p\,\bigl[\text{propagated mode }p\bigr]=\sum_{p=0}^{\infty}s^p\,g\,L_p(2\xi^2)\,e^{-2\mathrm{i} p\psi}.
\end{equation}
The left-hand side follows from linearity: the propagated sum is the sum of the propagated modes. Both sides are power series in $s$ that agree for every $s$ between $-1$ and $1$, so their coefficients agree one by one. Hence the mode with focal field $L_p(2\rho^2)e^{-\rho^2}$ propagates as $g\,L_p(2\xi^2)\,e^{-2\mathrm{i} p\psi}$: it keeps its shape in the co-moving coordinate $\xi$ and acquires the extra Gouy phase $e^{-2\mathrm{i} p\psi}$ in addition to that of $g$. Together with the factor $1/\nu=|\nu|^{-1}e^{-\mathrm{i}\psi}$ contained in $g$, the total Gouy factor of the mode is $e^{-\mathrm{i}(2p+1)\psi}$: the familiar Gouy phase $(2p+1)\psi$ of the mode $\mathrm{LG}_p^{0}$~\cite{Siegman1986}. Adding the modes with the weights $c_p$,
\begin{equation}
\frac{B}{g}=\sum_{p=0}^{n}c_p\,L_p(2\xi^2)\,e^{-2\mathrm{i} p\psi}.
\label{eq:s_B}
\end{equation}

\paragraph{Step 7. Summary of the propagation, and the reduced pair.} Equations~\eqref{eq:s_A} and \eqref{eq:s_B} together,
\begin{equation}
\frac{A}{g}=\xi^{\ell}e^{\mathrm{i}\ell\theta}e^{-\mathrm{i}\ell\psi},\qquad \frac{B}{g}=\sum_{p=0}^{n}c_p\,L_p(2\xi^2)\,e^{-2\mathrm{i} p\psi},
\label{eq:s_lift}
\end{equation}
are equation~(5) of the main text. The fields $A$ and $B$ obtained by multiplying these expressions by $g$ solve equation~\eqref{eq:s_parax} exactly and reduce to the prescribed focal fields at $z=z_0$, where $\psi=0$ and $\xi=\rho$. The plane-wave carrier $e^{\mathrm{i} kz}$ was already factored out in Step~1. The remaining common Gaussian factor $g$ is smooth and nonzero everywhere at finite $z$. Dividing both components by $g$ therefore leaves the ratio $A/B$, the polarization texture and all polarization preimages unchanged. It removes the common Gaussian amplitude, the wavefront-curvature phase and the fundamental Gouy phase, and leaves the reduced pair $(A/g,B/g)$ of equation~\eqref{eq:s_lift}. From now on we work with this reduced pair. Its normalized form (Step~9) is one particular smooth choice of the common phase of $A$ and $B$ at every point, a choice that the polarization does not see. It is convenient because its behavior at the boundary of space is controlled exactly. The proof below establishes the degree of this particular reduced field, and the corollary after Step~12 shows that the same integer is the linking number of closed polarization preimages, which depends on the polarization alone.

The coordinates $(\xi,\theta,\psi)$ describe all of space. Each value of $z$ corresponds to exactly one $\psi$ in the open interval $(-\pi/2,\pi/2)$, and in each plane each $r\ge0$ corresponds to exactly one $\xi\ge0$. All of space is therefore mapped one to one onto the slab $\{\xi\ge0,\ |\psi|<\pi/2\}$, with $\theta$ undefined only on the axis. In these coordinates, equation~\eqref{eq:s_lift} contains only polynomials in $\xi$ and the factors $e^{\mathrm{i}\ell\theta}$ and $e^{-\mathrm{i}\psi}$. It therefore extends continuously to the two faces $\psi=\pm\pi/2$, which represent the far field at $z=\pm\infty$, and its behavior far from the axis, $\xi\to\infty$, is decided by its highest powers of $\xi$. This is what makes the field at infinity easy to control.

\subsection*{Part 2: why the hopfion order is $n\ell$}

\paragraph{Step 8. The field at the boundary of space, and conditions (i) and (iii).} The boundary of space consists of three pieces.

(a) \emph{The far field}, $\psi\to\pm\pi/2$. There $e^{-2\mathrm{i} p\psi}\to e^{\mp\mathrm{i} p\pi}=(-1)^p$ and $e^{-\mathrm{i}\ell\psi}\to e^{\mp\mathrm{i}\ell\pi/2}=(\mp\mathrm{i})^{\ell}$, so that
\begin{equation}
\frac{A}{g}\to(\mp\mathrm{i})^{\ell}\,\xi^{\ell}e^{\mathrm{i}\ell\theta},\qquad \frac{B}{g}\to F(\xi^2)=\sum_{p=0}^{n}(-1)^pc_p\,L_p(2\xi^2).
\label{eq:s_far}
\end{equation}
$F$ is the far-field profile of the ring component; it is real, and it is the same on both faces. \emph{Condition (i)} requires $F(x)>0$ for all $x\ge0$. Then $B$ is real, positive and nonzero everywhere in the far field, on both sides of the focus.

(b) \emph{Far from the axis at a fixed $\psi$}, $\xi\to\infty$. By Step~T7 the highest power of $L_n(2\xi^2)$ is $(-1)^n2^n\xi^{2n}/n!$, and the modes with $p<n$ contain only lower powers. Hence
\begin{equation}
\frac{B}{g}=\frac{c_n(-2)^n}{n!}\,\xi^{2n}e^{-2\mathrm{i} n\psi}+(\text{terms with at most }\xi^{2n-2}),\qquad \Bigl|\frac{A}{g}\Bigr|=\xi^{\ell}.
\label{eq:s_large}
\end{equation}
In the same way, the highest power of $F(x)$ is $c_n2^nx^n/n!$; since $F$ must stay positive for large $x$, condition~(i) requires $c_n>0$. \emph{Condition (iii)} requires $\ell<2n$. Then $|A/B|\sim\xi^{\ell-2n}\to0$ as $\xi\to\infty$: far from the axis the $x$-polarized component $B$ dominates in every transverse plane, and the ratio tends to zero uniformly in $\psi$.

(c) \emph{The corners}, $\xi\to\infty$ and $\psi\to\pm\pi/2$ together. Both descriptions agree there: $e^{-2\mathrm{i} n\psi}\to(-1)^n$, so the leading term of equation~\eqref{eq:s_large} becomes $c_n2^n\xi^{2n}/n!>0$, the leading term of $F$.

\paragraph{Step 9. The normalized field and the values it takes on the boundary.} Following the main text, the reduced pair is normalized to
\begin{equation}
\mathbf u=\frac{(A/g,\ B/g)}{\bigl(|A/g|^2+|B/g|^2\bigr)^{1/2}},
\end{equation}
a pair of complex numbers whose squared moduli add up to one, that is, a point of $S^3$ (Step~T8). We use the same letters $A$ and $B$ for the two normalized components when we describe points of $S^3$. The Skyrme number, equation~(7) of the main text, is the signed volume of $S^3$ covered by $\mathbf u$ in units of $2\pi^2$. At each point, $\det[\mathbf u,\partial_1\mathbf u,\partial_2\mathbf u,\partial_3\mathbf u]\,\mathrm{d}^3q$ is the signed volume of the small piece of $S^3$ onto which a small coordinate box $\mathrm{d}^3q$ is mapped: the three vectors $\partial_i\mathbf u\,\mathrm{d}q_i$ span that small piece, and $\mathbf u$, the unit vector normal to $S^3$ at that point, fixes its sign. This holds in any coordinates $(q_1,q_2,q_3)$, including the co-moving coordinates $(\xi,\theta,\psi)$.

We now collect the values of $\mathbf u$ on the boundary of space found in Step~8.

On both far-field faces, $B/g=F(\xi^2)$ is real and positive, and the ratio $A/B$ equals $(\mp\mathrm{i})^{\ell}\xi^{\ell}e^{\mathrm{i}\ell\theta}$ divided by $F(\xi^2)$. Its modulus $R(\xi)=\xi^{\ell}/F(\xi^2)$ is continuous, equals 0 at $\xi=0$ and tends to 0 as $\xi\to\infty$ when $\ell<2n$. It therefore has a largest value $R_{\max}$ and takes every value between 0 and $R_{\max}$, while its phase $\ell\theta$ takes every value as $\theta$ goes around. The far-field values of $\mathbf u$ therefore fill the closed disc
\begin{equation}
\Delta=\Bigl\{\frac{(w,1)}{(1+|w|^2)^{1/2}}\ :\ |w|\le R_{\max}\Bigr\},
\end{equation}
the points of $S^3$ with $B$ real and positive and $|A/B|\le R_{\max}$. It is a two-dimensional surface in $S^3$, and both faces give the same disc.

Far from the axis, $A/B\to0$ and $B/g$ has the phase of $c_n(-1)^ne^{-2\mathrm{i} n\psi}$, which takes every value as $\psi$ varies. These values of $\mathbf u$ fill the circle
\begin{equation}
C=\{(0,e^{\mathrm{i}\gamma})\ :\ 0\le\gamma<2\pi\},
\end{equation}
the points of $S^3$ with $A=0$, which describe $x$-polarized light with every possible phase. At the corners, $\mathbf u\to(0,1)$, the single point shared by $\Delta$ and $C$.

The set of all boundary values is therefore $K=\Delta\cup C$, a disc and a circle that touch at one point. Two properties of $K$ are needed. First, $K$ has zero volume in $S^3$, because it consists of a surface and a curve. Second, removing $K$ does not cut $S^3$ into separate pieces: any two points of $S^3$ outside $K$ can be joined by a path that avoids $K$. The intuition is the same as in ordinary space, where a flat disc does not enclose any region (one can always walk around its rim), a circle does not either, and joining them at one point creates no enclosed pocket. The rigorous statement uses Alexander duality~\cite{Hatcher2002}. $K$ can be continuously contracted onto the circle $C$ (shrink the disc onto its point $(0,1)$), and as the union of a closed disc and a circle that meet at one point it is locally contractible, so its singular and \v{C}ech cohomology agree and Alexander duality applies. The reduced homology of $S^3\setminus K$ in degree zero, which counts the separate pieces of $S^3\setminus K$ minus one, equals the reduced cohomology of $K$ in degree two, which vanishes for a space that contracts onto a circle. Hence $S^3\setminus K$ is connected.

\paragraph{Step 10. The Skyrme number is an integer.} Add the boundary to the slab of Step~7: the two far-field faces and the limit $\xi\to\infty$. The result is a compact solid cylinder $D^2\times[-\pi/2,\pi/2]$, where the disc $D^2$ is the $(\xi,\theta)$ plane with the limit $\xi\to\infty$ added as its rim (for example, with the radial coordinate $\xi/(1+\xi)$, which runs from 0 on the axis to 1 at the rim). Inside it, $A/g$ vanishes only on the axis, and there $B/g=Q(e^{-2\mathrm{i}\psi})$; if $Q$ has no root on the unit circle $|t|=1$, the two components never vanish together, and $\mathbf u$ is defined and continuous everywhere inside. By Step~8, $\mathbf u$ then extends continuously to the whole cylinder and sends its boundary into $K$. Choose any point $y$ of $S^3$ outside $K$ at which the $3\times3$ Jacobian $J_{\mathbf u}$ of $\mathbf u$, taken in oriented local coordinates on the cylinder and on $S^3$, is invertible at every preimage (a regular value; by Sard's theorem, regular values fill all of $S^3$ except a set of zero volume). Then all preimages of $y$ lie inside the region, there are finitely many, and the signed count $\deg(y)$ of Step~T8 is well defined:
\begin{equation}
\deg(y)=\sum_{\mathbf x:\ \mathbf u(\mathbf x)=y}\mathrm{sign}\,\det J_{\mathbf u}(\mathbf x).
\end{equation}
By Step~T8, $\deg(y)$ can change only when $y$ crosses the set of boundary values $K$. By Step~9, any two such points $y$ can be joined by a path that avoids $K$. Hence $\deg(y)$ is the same integer, $\deg$, for almost every $y$. Since $K$ has zero volume, it does not contribute to the covered volume, and
\begin{equation}
N_S=\frac{1}{2\pi^2}\int\mathbf u^{*}\omega=\frac{1}{2\pi^2}\int_{S^3}\deg(y)\,\omega(y)=\deg ,
\end{equation}
where $\omega$ is the volume form of $S^3$ and $\mathbf u^{*}\omega$ is the integrand of equation~(7) of the main text. The Skyrme number is therefore an integer provided conditions (i) and (iii) hold and $Q(t)\neq0$ on $|t|=1$. The stronger condition (ii), which also excludes roots inside the unit circle, is used in Step~12 to show that this integer has magnitude exactly $n\ell$. To find its value, it is enough to count the preimages of one well-chosen point with their signs.

\paragraph{Step 11. Counting the preimages of the $y$-polarized state.} We choose the point $y=(A,B)=(e^{\mathrm{i}\chi},0)$ of $S^3$, with any fixed phase $\chi$. It lies above the $y$-polarized north pole of the Poincar\'e sphere: every $\chi$ describes the same $y$-polarized state, and fixing $\chi$ selects one value of the common phase. This is why each dark ring, which is entirely $y$-polarized, contains only a finite number of preimages of $y$. The point $y$ lies outside $K$: on the disc $\Delta$, $B>0$, and on the circle $C$, $|B|=1$. Its preimages are the points where
\begin{equation}
B=0\qquad\text{and}\qquad \arg(A/g)=\chi .
\end{equation}
\emph{Where is $B=0$?} By equation~\eqref{eq:s_B}, $B/g$ depends on $\xi$ and $\psi$ but not on $\theta$. Its zero set is therefore a set of circles around the axis, one circle for each zero $(\xi_j,\psi_j)$ of $B/g$ in the half-plane $H=\{\xi>0,\ |\psi|<\pi/2\}$: these circles are the dark rings. There are finitely many of them in a bounded part of $H$. None lies on the axis, because there $B/g=Q(t)\neq0$ (Steps~10 and 12); none lies close to the far-field faces, because there $B/g$ is close to $F>0$ by condition (i); and none lies at large $\xi$, because there the leading term of equation~\eqref{eq:s_large} dominates (Step~T6).

\emph{Where on a ring is $\arg(A/g)=\chi$?} On the $j$-th ring, equation~\eqref{eq:s_A} gives $A/g=\xi_j^{\ell}e^{\mathrm{i}\ell(\theta-\psi_j)}$, which is nonzero because $\xi_j>0$. Its phase $\ell(\theta-\psi_j)$ increases by $2\pi\ell$ once around the ring, so it equals $\chi$ (up to multiples of $2\pi$) at exactly $\ell$ values of $\theta$. Each ring therefore contains exactly $\ell$ preimages of $y$.

\emph{The sign of each preimage.} Near $y$, a point of $S^3$ can be described by three real coordinates: the phase $\alpha=\arg A$ and the real and imaginary parts of $B'=B/|A|$, which are small near $y$. In terms of the field, $\alpha=\ell(\theta-\psi)$ and $B'=(B/g)/\xi^{\ell}$. The derivative matrix of the map $(\theta,\xi,\psi)\mapsto(\alpha,\mathrm{Re}B',\mathrm{Im}B')$ at a preimage is
\begin{equation}
\begin{pmatrix}\ell & \partial_\xi\alpha & \partial_\psi\alpha\\ 0 & \partial_\xi\,\mathrm{Re}B' & \partial_\psi\,\mathrm{Re}B'\\ 0 & \partial_\xi\,\mathrm{Im}B' & \partial_\psi\,\mathrm{Im}B'\end{pmatrix},
\end{equation}
where the zeros in the first column express that $B'$ does not depend on $\theta$. Its determinant is $\ell$ times the determinant of the lower-right $2\times2$ block. At a zero of $B/g$, the derivatives of $B'$ are those of $B/g$ divided by the positive number $\xi_j^{\ell}$, so by Step~T3 the sign of this block is the index $\varepsilon_j=\pm1$ of the zero of $B/g$ at $(\xi_j,\psi_j)$. The change from $(x,y,z)$ to $(\theta,\xi,\psi)$ and the choice of coordinates on $S^3$ contribute one further sign $\sigma_0$, which is the same at every point. The sign of the preimage is therefore $\sigma_0\varepsilon_j$, and all $\ell$ preimages on the $j$-th ring carry the same sign. If all zeros of $B/g$ are simple, $y$ is a regular value, and adding up gives
\begin{equation}
\deg=\sigma_0\,\ell\sum_j\varepsilon_j .
\label{eq:s_deg}
\end{equation}
If a zero of $B/g$ is not simple, take instead a nearby regular point $y'$ outside $K$. The preimages of $y'$ that lie near that zero have a total signed contribution $\sigma_0\ell$ times the index of the zero (Step~T3). Since $\deg(y)$ has the same value for all points outside $K$ (Step~10), equation~\eqref{eq:s_deg} remains valid with every zero counted by its index. The counting assumes that the zeros of $B/g$ in $H$ are isolated, as they are for generic coefficients. If they were not, a small change of the coefficients $c_p$ that keeps conditions (i)--(iii) would make them isolated and simple; the integer $\deg$ cannot change under such a change, because the boundary values stay away from $y$ throughout, so the result is the same.

\paragraph{Step 12. The algebraic ring index, and condition (ii).} It remains to find $\sum_j\varepsilon_j$, the sum of the indices of all zeros of $B/g$ in $H$. By Step~T4 it equals the number of turns of $B/g$ along the boundary of a rectangle that contains all zeros, traversed counterclockwise. We use the rectangle $0\le\xi\le X$, $-\pi/2\le\psi\le\pi/2$ in the plane with $\xi$ horizontal and $\psi$ vertical, with $X$ so large that all zeros lie inside. Its four sides contribute as follows.
\begin{enumerate}
\item Bottom side, $\psi=-\pi/2$, $\xi$ from 0 to $X$: $B/g=F(\xi^2)$ is real and positive by condition (i), so its argument stays 0. Contribution: 0.
\item Right side, $\xi=X$, $\psi$ from $-\pi/2$ to $\pi/2$: by equation~\eqref{eq:s_large}, $B/g$ is the leading term $f_0=c_n(-2)^nX^{2n}e^{-2\mathrm{i} n\psi}/n!$ plus terms that are smaller for large $X$. The argument of $f_0$ is $-2n\psi$ plus a constant, which decreases by $2\pi n$ as $\psi$ increases by $\pi$: $f_0$ turns $n$ times clockwise. At both ends of this side, $B/g=F(X^2)$ and $f_0=c_n2^nX^{2n}/n!$ are real and positive, so by Step~T6 $B/g$ turns exactly as $f_0$ does. Contribution: $-n$.
\item Top side, $\psi=\pi/2$, $\xi$ from $X$ to 0: as for the bottom side. Contribution: 0.
\item Left side, on the axis, $\xi=0$, $\psi$ from $\pi/2$ to $-\pi/2$: since $L_p(0)=1$,
\begin{equation}
\frac{B}{g}=\sum_{p=0}^{n}c_p\,e^{-2\mathrm{i} p\psi}=Q(t),\qquad t=e^{-2\mathrm{i}\psi}.
\end{equation}
As $\psi$ decreases from $\pi/2$ to $-\pi/2$, the angle $-2\psi$ increases from $-\pi$ to $\pi$, so $t$ runs once counterclockwise around the unit circle, from $-1$ back to $-1$. By Step~T5, $Q$ then makes $m_0$ turns, where $m_0$ is the number of roots of $Q$ with $|t|<1$, counted with multiplicity. \emph{Condition (ii)} requires that $Q$ has no roots with $|t|\le1$. It excludes roots on the circle, where $B$ would vanish on the axis together with $A$, and it gives $m_0=0$. Contribution: $m_0$, which is 0 under condition (ii).
\end{enumerate}
Adding the four sides,
\begin{equation}
\sum_j\varepsilon_j=m_0-n=-n .
\end{equation}
The zero set of $B$ therefore consists of rings whose indices add up to $-n$; this sum is the algebraic ring index. Additional zeros of opposite index can occur only in pairs whose indices cancel. With equation~\eqref{eq:s_deg},
\begin{equation}
|N_S|=|\deg|=n\ell ,
\end{equation}
which is equation~(8) of the main text. In words: the $y$-polarized preimage is the union of the dark rings, whereas the $x$-polarized preimage is the beam axis. The phase of $A$ winds with magnitude $\ell$ along each ring, because the axis is a vortex line of charge $\ell$, with the orientation set by the ring index, so the magnitude of the total signed phase winding along the rings is $n\ell$. The corollary below shows that the same integer is the linking number of the closed preimages of any two states that the far field never reaches. For all designs used in this work, a scan of $H$ on a $400\times400$ grid finds exactly $n$ zeros, each of index $-1$, all at $\psi=0$ to within the grid step.

\paragraph{Corollary: closed preimages and their linking.} Steps~10--12 fix the degree of the normalized pair $\mathbf u$. We now show that this integer is also a linking number of polarization preimages, which depends on the texture $\mathbf n$ alone. Write $h$ for the Hopf map, which sends a point $(A,B)$ of $S^3$ to its polarization state $w=A/B$ on the Poincar\'e sphere, so that $\mathbf n=h\circ\mathbf u$. By Step~9, $h$ maps the disc $\Delta$ onto the closed cap $E=\{|w|\le R_{\max}\}$ around the $x$-polarized south pole and the circle $C$ onto the south pole itself. All boundary values of $\mathbf n$ therefore lie in $E$.

\emph{Statement.} Let $s_1\neq s_2$ be two states with $|w_1|,|w_2|>R_{\max}$ that are regular values of $\mathbf n$. Then their preimages are finite unions of closed loops inside the beam, and their total linking number satisfies $|Lk(\mathbf n^{-1}(s_1),\mathbf n^{-1}(s_2))|=|N_S|=n\ell$.

\emph{(a) The preimages are closed.} $\mathbf n^{-1}(s_i)$ is a compact curve in the cylinder of Step~10 (a closed subset of a compact set). It cannot reach the boundary, where $\mathbf n$ lies in $E$, which does not contain $s_i$. Near each of its points it is a smooth curve without end points, because $s_i$ is a regular value. It is therefore a finite union of closed loops.

\emph{(b) Closing the texture without touching the two preimages.} Choose $R'$ with $R_{\max}<R'<\min(|w_1|,|w_2|)$. Because $\mathbf n$ is continuous on the compact cylinder and its boundary values satisfy $|w|\le R_{\max}$, there is a thin shell $U$ along the boundary on which $|w|<R'$. Label the points of $U$ as $(b,\tau)$, with $b$ on the boundary and the depth $\sigma\in[0,1]$, $\sigma=0$ on the boundary and $\sigma=1$ on the inner face of the shell. Define a new texture $\mathbf n'$ that equals $\mathbf n$ outside $U$ and on $U$ is the state with ratio $\sigma\,w(b,\sigma)$. It is continuous, equals $\mathbf n$ on the inner face, takes only states with $|w|<R'$ on $U$, and is the $x$-polarized state on the whole boundary. Since the cylinder is a closed ball, collapsing its boundary to one point gives a three-sphere, and $\mathbf n'$ is a map of this three-sphere to the Poincar\'e sphere. Its preimages of $s_1$ and $s_2$ are those of $\mathbf n$, because $\mathbf n'$ does not reach $s_1$ or $s_2$ on $U$. ($\mathbf n'$ may be smoothed on $U$ without leaving the cap $|w|<R'$, so it can be taken smooth.)

\emph{(c) Hopf invariant.} For a smooth map $f$ of $S^3$ to $S^2$, the linking number of the preimages of two distinct regular values is the same for every such pair; it is the Hopf invariant $H(f)$~\cite{Hopf1931,BottTu1982}. Hence $Lk(\mathbf n^{-1}(s_1),\mathbf n^{-1}(s_2))=H(\mathbf n')$.

\emph{(d) Hopf invariant and degree.} Every map of $S^3$ to $S^2$ lifts through the Hopf map: the circle bundle that $\mathbf n'$ pulls back from the Hopf fibration is trivial, because the second cohomology of $S^3$ vanishes, so it has a section. There is therefore a map $\mathbf u'$ of $S^3$ to $S^3$ with $h\circ\mathbf u'=\mathbf n'$, and for such a composition $H(h\circ\mathbf u')=\deg(\mathbf u')\,H(h)$ with $H(h)=\pm1$~\cite{BottTu1982}. Hence $|H(\mathbf n')|=|\deg\mathbf u'|$.

\emph{(e) The lift has the degree of $\mathbf u$.} Outside $U$, $\mathbf u$ and $\mathbf u'$ are lifts of the same texture, so $\mathbf u'=e^{\mathrm{i}\chi}\mathbf u$ with a continuous common phase factor $e^{\mathrm{i}\chi}$. The region outside $U$ is a solid cylinder, in which every closed loop can be contracted, so $\chi$ can be chosen as a single-valued real function there. Take a point $y$ of $S^3$ above $s_1$ that is a regular value of both $\mathbf u$ and $\mathbf u'$. Its preimages lie on the loops of (a), at the points where the common phase of the pair takes the value fixed by $y$. On each oriented loop, the signed number of these points is the number of turns of the common phase along the loop (Step~T2). Multiplying by $e^{\mathrm{i}\chi}$ adds the number of turns of $e^{\mathrm{i}\chi}$ along the loop, which is zero because $\chi$ is single valued. Hence $\deg\mathbf u'=\deg\mathbf u=N_S$, and with (c), (d) and Step~12,
\begin{equation}
|Lk(\mathbf n^{-1}(s_1),\mathbf n^{-1}(s_2))|=|N_S|=n\ell .
\end{equation}

\emph{Which states close.} $R_{\max}=\max_{x\ge0}x^{\ell/2}/F(x)$ is finite under conditions (i) and (iii). A positive rescaling of $B$ divides it by the same factor and changes neither the degree nor the conclusion. With the components scaled to equal power in the focal plane, $R_{\max}$ lies between 1.1 and 2.3 for every design of this work (Supplementary Table~\ref{tab:designs}). The states $|w|=8$ and 30 traced in this work therefore have closed preimages in every ideal target. Close to the $y$-polarized pole these preimages lie on thin tori around the dark rings, one loop per ring, and each winds $\ell$ times around its ring (main-text Fig.~3a), which is the picture behind the count of Step~11.

\paragraph{Existence of admissible designs of every order.} The three conditions can be satisfied for every positive integer $n$, so equation~(8) of the main text is not an empty statement for any order. Set $c_n=1$, $c_p=0$ for $0<p<n$, and $c_0=C>0$. Then
\begin{equation}
F(x)=C+(-1)^nL_n(2x),\qquad Q(t)=C+t^n .
\end{equation}
The polynomial $(-1)^nL_n(2x)$ has the positive leading coefficient $2^n/n!$ (Step~T7), so it tends to $+\infty$ as $x\to\infty$ and has a finite minimum $M_n$ on $x\ge0$. Choosing $C>|M_n|$ makes $F(x)>0$ for all $x\ge0$, which is condition (i). The $n$ roots of $Q(t)=C+t^n$ all have modulus $C^{1/n}$, so choosing also $C>1$ places all of them outside the unit circle, which is condition (ii). Any $\ell<2n$, for example $\ell=1$, satisfies condition (iii), and equation~(8) of the main text gives $|N_S|=n\ell$. An admissible ideal target therefore exists for every positive integer order $n\ell$. The particular coefficients used in this work, $(3,2,2)$ and $(4,3,2,2)$, are chosen not for existence but to place the rings as far as possible into the bright core of the Gaussian envelope, where a phase-only sheet can realize them (Step~15 and Supplementary Note~\ref{sn:general}).

\paragraph{Step 13. What happens when a condition fails.} Each condition protects one part of the argument.

\emph{Condition (ii).} If $Q$ has a root on the unit circle, $t_*=e^{-2\mathrm{i}\psi_*}$, then $B$ vanishes on the axis in the plane $\psi=\psi_*$, where $A$ also vanishes. The total field is zero at that point, the polarization is undefined there, and a dark ring touches the axis. For example, $(c_0,c_1,c_2)=(1,1,1)$ gives $Q(t)=1+t+t^2$ with roots $e^{\pm2\pi\mathrm{i}/3}$ on the unit circle; its two rings exist, but the field vanishes on the axis at $\psi=\mp\pi/3$, and $|N_S|$ evaluated over all space stays between 1.1 and 1.3 instead of approaching 2. If conditions (i) and (iii) remain satisfied, no root lies on $|t|=1$, and $Q$ has $m_0$ roots inside the unit circle, counted with multiplicity, the count of Step~12 gives $|N_S|=|n-m_0|\,\ell$.

\emph{Condition (i).} If $F$ has a root $x_*>0$, then $B$ vanishes on a far-field face, on the cone $\xi^2=x_*$, where $A\neq0$: a dark ring escapes to infinity, the far field reaches the $y$-polarized state, $y$ lies in the set of boundary values, and the count is no longer defined. If $F(0)=0$, then $A$ and $B$ both vanish on the axis in the far field, and the normalized field is undefined there. Either case invalidates the boundary argument of Steps~9 and 10.

\emph{Condition (iii).} For $\ell=2n$, $|A/B|$ tends to the constant $n!/(c_n2^n)$ as $\xi\to\infty$, while the phases of $A$ and $B$ vary independently with $\theta$ and $\psi$. The boundary is then mapped onto a torus in $S^3$ (the points with this fixed value of $|A/B|$), which divides $S^3$ into two regions with different counts $\deg(y)$. A single degree is then not defined, and $N_S$, a weighted average of the two counts, need not be an integer. Its value can nevertheless be computed exactly. Write $\eta=|A/g|^2/(|A/g|^2+|B/g|^2)$ and $\beta=\arg(B/g)$. The volume form of $S^3$ is $\tfrac12\,\mathrm{d}\eta\wedge\mathrm{d}\alpha\wedge\mathrm{d}\beta$ with $\alpha=\arg A$, and $\alpha=\ell(\theta-\psi)$, while $\eta$ and $\beta$ do not depend on $\theta$. Integrating over $\theta$ gives, up to the orientation sign, $N_S=(\ell/2\pi)\int_H\mathrm{d}\eta\wedge\mathrm{d}\beta$. By Stokes' theorem this equals $(\ell/2\pi)$ times the integral of $\eta\,\mathrm{d}\beta$ around the rectangle of Step~12 minus small circles around the zeros of $B/g$. On the far-field faces $\beta$ is constant, on the axis $\eta=0$, around each zero $\eta=1$, and far from the axis $\eta$ tends to a constant $\eta_\infty$ while $\beta$ makes $n$ clockwise turns. Hence
\begin{equation}
|N_S|=\ell\,\bigl|\,n-m_0-n\,\eta_\infty\bigr| .
\end{equation}
For $\ell<2n$, $\eta_\infty=0$ and this is the result of Step~12. For $\ell=2n$, equation~\eqref{eq:s_large} gives $\eta_\infty=1/(1+b^2)$ with $b=2^nc_n/n!$, so that, for $m_0=0$,
\begin{equation}
|N_S|=n\ell\,\frac{b^2}{1+b^2}.
\end{equation}
This value depends on the relative amplitude of $A$ and $B$, which confirms that it is not topological. For the single ring with $\ell=2$, the device coefficients $(c_0,c_1)=(\tfrac34,\tfrac14)$ give $b=\tfrac12$ and $|N_S|=\tfrac25$, whereas the same ring scaled to $(c_0,c_1)=(1,\tfrac13)$ gives $|N_S|=\tfrac{8}{13}=0.615$. For $\ell>2n$, $A$ dominates at large $\xi$, and the $y$-polarized states themselves lie in the set of boundary values.

\subsection*{Part 3: examples and the two phase maps}

\paragraph{Step 14. The single-ring family.} For $n=1$, the coefficients $c_0=1-\kappa/2$ and $c_1=\kappa/2$ with $\kappa>0$ give, with $L_1(y)=1-y$,
\begin{equation}
B_f=\Bigl[1-\frac{\kappa}{2}+\frac{\kappa}{2}(1-2\rho^2)\Bigr]e^{-\rho^2}=(1-\kappa\rho^2)\,e^{-\rho^2},
\end{equation}
with its dark ring at $\rho^2=1/\kappa$, that is, at $r_c=w_0/\sqrt\kappa$. Equation~\eqref{eq:s_B} becomes
\begin{equation}
\frac{B}{g}=1-\frac{\kappa}{2}+\Bigl(\frac{\kappa}{2}-\kappa\xi^2\Bigr)e^{-2\mathrm{i}\psi},\qquad
\frac{B}{g}\,\nu^2=\bigl[1-(1-\kappa)\tau^2-\kappa\rho^2\bigr]+\mathrm{i}\,\tau(2-\kappa).
\label{eq:s_single}
\end{equation}
The second form follows from the first after multiplication by $\nu^2=1-\tau^2+2\mathrm{i}\tau$, using $\nu^2e^{-2\mathrm{i}\psi}=\nu\nu^{*}=1+\tau^2$ (identities I2 and I3) and $\xi^2(1+\tau^2)=\rho^2$. Since $\nu^2\neq0$, $B$ vanishes exactly where the second form vanishes. Its imaginary part vanishes only at $\tau=0$ (for $\kappa\neq2$), that is, in the focal plane, and there the real part $1-\kappa\rho^2$ vanishes only at $\rho^2=1/\kappa$. The zero set of $B$ is therefore a single ring in the focal plane.

The three conditions become simple inequalities:
\begin{itemize}
\item Condition (i): $F(x)=c_0-c_1L_1(2x)=1-\frac{\kappa}{2}-\frac{\kappa}{2}(1-2x)=1-\kappa+\kappa x$, which is positive for all $x\ge0$ if and only if $\kappa<1$.
\item Condition (ii): $Q(t)=1-\frac{\kappa}{2}+\frac{\kappa}{2}t$ has its single root at $t=-(2-\kappa)/\kappa$. Its modulus is $|t|=|2-\kappa|/\kappa$, which exceeds one exactly for $0<\kappa<1$.
\item Condition (iii): $\ell<2n=2$ permits $\ell=1$.
\end{itemize}
Conditions (i) and (ii) therefore both reduce to $0<\kappa<1$. For $\ell=1$, the far-field ratio is $|A/B|=\xi/(1-\kappa+\kappa\xi^2)$. Setting its derivative to zero, $1-\kappa-\kappa\xi^2=0$, shows that it reaches its maximum $[2\sqrt{\kappa(1-\kappa)}]^{-1}$ at $\xi^2=(1-\kappa)/\kappa$; for the device, $\kappa=1/2$, this maximum is 1. At $\kappa=1$, $F(0)=0$ and both components vanish on the axis at $z=\pm\infty$; for $\kappa>1$, $F$ vanishes at $\xi^2=(\kappa-1)/\kappa$, so that $B$ vanishes on a cone in the far field. In both cases the far field reaches every polarization state.

\paragraph{Step 15. Two and three rings.} For $n=2$ we use $(c_0,c_1,c_2)=(3,2,2)$. With $x=\rho^2$, $L_1(2x)=1-2x$ and $L_2(2x)=1-4x+2x^2$:
\begin{equation}
\frac{B_f}{e^{-\rho^2}}=3+2(1-2x)+2(1-4x+2x^2)=7-12x+4x^2 ,
\end{equation}
\begin{equation}
F(x)=3-2(1-2x)+2(1-4x+2x^2)=4x^2-4x+3,\qquad Q(t)=3+2t+2t^2 .
\label{eq:s_two}
\end{equation}
The focal polynomial has the two roots $x=(12\mp\sqrt{144-112})/8=(3\mp\sqrt2)/2=0.793$ and 2.207, so the dark rings lie at $r=0.891\,w_0$ and $1.486\,w_0$ in the focal plane. $F$ has no real roots, since its discriminant is $16-48<0$, and it is positive at $x=0$; so condition (i) holds. The roots of $Q$ are $t=(-2\pm\sqrt{4-24})/4=(-1\pm\mathrm{i}\sqrt5)/2$, with modulus $\sqrt{(1+5)/4}=\sqrt{3/2}>1$; so condition (ii) holds. Condition (iii) allows $\ell=1$, 2 and 3. The hopfion orders are therefore 2, 4 and 6. Both rings lie in the bright part of the beam (Supplementary Table~\ref{tab:designs}).

For $n=3$ we use $(c_0,c_1,c_2,c_3)=(4,3,2,2)$, the three-ring analogue of $(3,2,2)$. With $L_3(2x)=1-6x+6x^2-\tfrac43x^3$,
\begin{equation}
\frac{B_f}{e^{-\rho^2}}=11-26x+16x^2-\tfrac83x^3,\qquad F(x)=1+10x-8x^2+\tfrac83x^3,\qquad Q(t)=4+3t+2t^2+2t^3 .
\end{equation}
The focal polynomial has three positive roots, $x=0.665$, 1.712 and 3.623, so the dark rings lie at $r=0.816\,w_0$, $1.308\,w_0$ and $1.903\,w_0$. The derivative $F'(x)=10-16x+8x^2$ has a negative discriminant, $256-320<0$, so $F$ increases monotonically from $F(0)=1$ and is positive for all $x\ge0$; condition (i) holds. The roots of $Q$ are $t=-1.174$ and $0.087\pm1.302\,\mathrm{i}$, with moduli 1.174 and 1.305; condition (ii) holds. Condition (iii) allows $\ell=1,\dots,5$, that is, orders 3, 6, 9, 12 and 15. The vortex order also sets how bright the rings are, because $|A|^2\propto\rho^{2\ell}e^{-2\rho^2}$ is small near the axis for large $\ell$ and small far out for small $\ell$; $\ell=3$ balances the two (Supplementary Table~\ref{tab:designs}).

\begin{table}[!htbp]
\centering
\caption{\textbf{Ring components of this work.} Rings: focal-plane positions $\rho^2$. $|t|_{\min}$: smallest root modulus of $Q(t)$. $R_{\max}$: largest $|w|$ reached in the far field, and ring intensities: $|A|^2$ at each ring in percent of the focal-plane maximum of $|A|^2+|B|^2$, both with the components scaled to equal focal power.}
\label{tab:designs}
\small
\begin{tabular}{lccccccc}
\toprule
$(c_0,\dots,c_n)$ & $F(x)$ & rings $\rho^2$ & $|t|_{\min}$ & $\ell$ & $n\ell$ & $R_{\max}$ & ring intensities (\%)\\
\midrule
$(\tfrac34,\tfrac14)$ & $\tfrac12(1+x)$ & 2 & 3 & 1 & 1 & 1.12 & 4.6\\
\midrule
$(3,2,2)$ & $4x^2-4x+3$ & 0.793, 2.207 & 1.225 & 1 & 2 & 2.26 & 11.3, 1.9\\
& & & & 2 & 4 & 1.99 & 8.9, 4.1\\
& & & & 3 & 6 & 1.60 & 4.7, 6.0\\
\midrule
$(4,3,2,2)$ & $1+10x-8x^2+\tfrac83x^3$ & 0.665, 1.712, 3.623 & 1.174 & 1 & 3 & 1.46 & 9.6, 3.0, 0.14\\
& & & & 2 & 6 & 1.75 & 6.4, 5.2, 0.51\\
& & & & 3 & 9 & 1.86 & 2.8, 6.0, 1.2\\
& & & & 4 & 12 & 1.82 & 0.94, 5.1, 2.2\\
& & & & 5 & 15 & 1.69 & 0.25, 3.5, 3.2\\
\bottomrule
\end{tabular}
\end{table}

\paragraph{Step 16. The two phase maps.} The sheet lies at $z=0$, a distance $z_0$ before the focus. There $\tau=-z_0/z_R=-\zeta$, so that
\begin{equation}
\nu_s=1-\mathrm{i}\zeta,\qquad |\nu_s|^2=1+\zeta^2,\qquad \psi=-\arctan\zeta,\qquad \xi^2=\frac{\rho^2}{1+\zeta^2},
\end{equation}
where the subscript $s$ denotes the sheet. From equation~\eqref{eq:s_g}, with $1/\nu_s=(1+\mathrm{i}\zeta)/(1+\zeta^2)$,
\begin{equation}
g=\frac{1+\mathrm{i}\zeta}{1+\zeta^2}\exp\Bigl[-\frac{\rho^2(1+\mathrm{i}\zeta)}{1+\zeta^2}\Bigr],\qquad \arg g=\arctan\zeta-\frac{\zeta\rho^2}{1+\zeta^2}.
\end{equation}
The phase of a product is the sum of the phases (Step~T1), so $\arg A=\arg g+\arg(A/g)$ and $\arg B=\arg g+\arg(B/g)$. With $-\ell\psi=\ell\arctan\zeta$ and $e^{-2\mathrm{i} p\psi}=e^{2\mathrm{i} p\arctan\zeta}$, equation~\eqref{eq:s_lift} gives
\begin{equation}
\begin{aligned}
\phi_y&=\arg A=\ell\theta+(\ell+1)\arctan\zeta-\frac{\zeta\rho^2}{1+\zeta^2},\\
\phi_x&=\arg B=\arg\Bigl[\sum_{p=0}^{n}c_p\,L_p\Bigl(\frac{2\rho^2}{1+\zeta^2}\Bigr)e^{2\mathrm{i} p\arctan\zeta}\Bigr]+\arctan\zeta-\frac{\zeta\rho^2}{1+\zeta^2},
\end{aligned}
\end{equation}
which is equation~(9) of the main text. The last term is common to both channels. Since $\zeta\rho^2/(1+\zeta^2)=kr^2/(2R)$ with $R=z_R(1+\zeta^2)/\zeta=z_0(1+\zeta^{-2})$, it is the phase $-kr^2/(2R)$ of a converging lens of focal length $R$. For the single ring, the second form of equation~\eqref{eq:s_single} with $\tau=-\zeta$ and $\arg\nu_s^2=-2\arctan\zeta$ gives the radial term $\arg\{[1-(1-\kappa)\zeta^2-\kappa\rho^2]+\mathrm{i}\,\zeta(\kappa-2)\}+2\arctan\zeta$. For the device, $w_0=1.5\lambda$ and $z_0=10\lambda$ give $z_R=\pi w_0^2/\lambda=7.07\lambda$, $\zeta=1.415$, $\arctan\zeta=0.955$~rad ($\ang{54.7}$) and $R=15.0\lambda$.

\paragraph{Step 17. Design procedure.} (1) Choose the radial order $n$ and coefficients $c_0,\dots,c_n$ such that the far-field profile $F(x)$ of equation~\eqref{eq:s_far} is positive for $x\ge0$ and the polynomial $Q(t)=\sum_pc_pt^p$ has all its roots outside the unit circle. (2) Choose a vortex order $\ell<2n$; the hopfion order is $n\ell$. (3) Choose $w_0$ and $z_0$ and evaluate the two phases of Step~16 at the cell centers. No step involves an iterative or numerical optimization.

\section{The meta-atom}
\label{sn:cell}
The unit cells were simulated in CST Studio Suite with unit-cell boundaries and Floquet ports, with gold as a lossy metal and fused quartz of $\epsilon_r=3.75$; the geometry is shown in Fig.~4 of the main text. Each phase is set by its own arm, the orthogonal arm barely perturbs it, and the phase changes by less than \ang{2.5} up to \ang{5} incidence (Supplementary Table~\ref{tab:uc}, Supplementary Fig.~\ref{fig:uc11}). Each value of equation~(9) of the main text is assigned the library cell of nearest phase.

\begin{table}[!htbp]
\centering
\caption{\textbf{Unit-cell library and mapping} (\SI{1}{\tera\hertz}).}
\label{tab:uc}
\small
\begin{tabular}{>{\raggedright\arraybackslash}p{0.55\linewidth}>{\raggedright\arraybackslash}p{0.38\linewidth}}
\toprule
Quantity & Value\\
\midrule
Arm length of the library (other arm \SI{85}{\micro\meter}) & 57--\SI{107.5}{\micro\meter}\\
Phase range; amplitude & \ang{360}; 0.71--0.94\\
Phase change caused by the orthogonal arm & \ang{1.6} r.m.s., at most \ang{5.9}\\
Cross-polarized transmission & below \SI{-38}{\decibel}\\
Phase change at \ang{5} incidence & below \ang{2.5}\\
Phase error of the selected cells & \ang{1.8} r.m.s.\\
Mean amplitude of the selected cells & 0.87\\
\bottomrule
\end{tabular}
\end{table}

\sfig{SI_unitcell_11x11.jpg}{1}{\textbf{Two-dimensional unit-cell sweep at \SI{1}{\tera\hertz}} ($11\times11$ values of $L_x$ and $L_y$). \textbf{a},\,\textbf{d},~Transmission phases $\phi_x$ and $\phi_y$. \textbf{b},\,\textbf{e},~Amplitudes. \textbf{c},\,\textbf{f},~Cross-arm residuals $\Delta\phi_x$ and $\Delta\phi_y$ (deviation from the mean over the orthogonal arm).}{fig:uc11}

\paragraph{Fabrication.} No device has been fabricated in this work. The stack is Au / \SI{15}{\micro\meter} fused quartz / Au / \SI{15}{\micro\meter} / Au / \SI{15}{\micro\meter} / Au, with four \SI{0.3}{\micro\meter} gold planes and a \SI{4.2}{\milli\meter} aperture (Supplementary Fig.~\ref{fig:fab}). Plasma-activated direct bonding of fused silica has been reported at about 200--\SI{250}{\celsius}~\cite{Kalkowski2012}. Direct bonding requires sufficiently smooth and planar surfaces~\cite{Moriceau2010}; we therefore propose recessing and planarizing each internal gold plane before bonding. The stack is built on a carrier: each quartz layer is a bonded full-thickness wafer thinned in place, the internal gold planes are patterned into shallow recesses and planarized, the top plane is patterned on the free surface, and the bottom plane is patterned after release, aligned through the glass. The smallest feature ($L_w=\SI{10}{\micro\meter}$) is compatible with UV lithography, and multilayer terahertz metamaterials with patterned metal layers in dielectric stacks have been demonstrated at comparable frequencies~\cite{Tao2008,Grady2013}, although with different spacer materials and processes. The dielectric properties of the actual wafers, measured by terahertz time-domain spectroscopy~\cite{Naftaly2007}, should be used to regenerate the library before mapping. The device also needs an aperture that confines the illumination to the sheet (Supplementary Note~\ref{sn:range}). A Monte Carlo study of the main fabrication deviations keeps the first-order hopfion in every realization (Supplementary Note~\ref{sn:range}).

\sfig{SI_fab_flow.jpg}{0.9}{\textbf{Proposed fabrication route} (cross-sections). The internal gold planes are recessed and planarized before each low-temperature bond, and each \SI{15}{\micro\meter} layer is obtained by thinning a bonded wafer on a carrier.}{fig:fab}

\section{The first-order device}
\label{sn:first}
Supplementary Table~\ref{tab:params} lists the parameters of all simulations. Supplementary Figs.~\ref{fig:mapsc} and \ref{fig:mapfw} show the fields of the first-order device in the scalar model and in the full-wave simulation; in both, $E_y$ carries the axial vortex, $E_x$ forms the dark ring, and the polarization winds once around the axis. Supplementary Tables~\ref{tab:first} and \ref{tab:freq} collect the topological measures. The ring and three generic preimages are traced directly; the Skyrme number and the Whitehead integral are integral diagnostics, evaluated on a modal reconstruction, in which the field on one transverse plane is projected onto Laguerre--Gauss modes and propagated in closed form over all space, with a weak incoherent $x$-polarized background of relative level $\varepsilon$,
\begin{equation}
Q_W=\frac{1}{(4\pi)^2}\int\mathbf F\cdot\mathbf a\,\mathrm{d}^3q,\qquad F_i=\tfrac12\epsilon_{ijk}\,\mathbf n_\varepsilon\cdot(\partial_j\mathbf n_\varepsilon\times\partial_k\mathbf n_\varepsilon),\qquad \nabla\times\mathbf a=\mathbf F,
\end{equation}
with $\mathbf n_\varepsilon$ the texture after the background is added~\cite{Whitehead1947}. They depend on the background and serve only as a cross-check of the traced loops. The full-wave field departs from the scalar model in its waveform, yet keeps its topology; above the design frequency the strong dispersion of the cell library near \SI{1.04}{\tera\hertz} begins to degrade it.

\begin{table}[!htbp]
\centering
\caption{\textbf{Parameters of the simulations.}}
\label{tab:params}
\small
\begin{tabular}{>{\raggedright\arraybackslash}p{0.28\linewidth}>{\raggedright\arraybackslash}p{0.64\linewidth}}
\toprule
Design & $\lambda=\SI{300}{\micro\meter}$ (\SI{1}{\tera\hertz}); $w_0=1.5\lambda$; $z_0=10\lambda$ ($z_R=7.07\lambda$, $\zeta=1.415$); $28\times28$ cells, pitch $\lambda/2$\\
Illumination & Gaussian of waist $6\lambda$, flat phase, \ang{45} polarization; cosine taper to zero between $r=6.5\lambda$ and $7\lambda$\\
Scalar model & $4\times4$ samples per cell; window four times the sheet; volume $|x|,|y|\le7\lambda$, $0<z\le20\lambda$, axial step $0.1\lambda$\\
Ideal Skyrme number & grids of $101^3$ to $401^3$ points; $h^2$ extrapolation from the three finest\\
Loops and morph & 120 meridional half-planes; loops Fourier-interpolated 16-fold; 41 steps per segment of the morph; rings traced within $r\le3.5w_0$\\
Full-wave model & gold, lossy metal, $\sigma=4.561\times10^{7}$~S\,m$^{-1}$; fused quartz, $\epsilon_r=3.75$, $\tan\delta=4\times10^{-4}$; open boundaries\\
Full-wave source and export & source plane $z=\SI{-0.6}{\milli\meter}$, numerical aperture below 0.9; field exported at 0.975, 1.000 and \SI{1.025}{\tera\hertz} on a \SI{25}{\micro\meter} grid, $|x|,|y|\le\SI{3.06}{\milli\meter}$, $-0.71\le z\le\SI{6.09}{\milli\meter}$; analysis from $z=\SI{0.3}{\milli\meter}$\\
\bottomrule
\end{tabular}
\end{table}

\begin{table}[!htbp]
\centering
\caption{\textbf{First-order device at \SI{1}{\tera\hertz}.}}
\label{tab:first}
\small
\begin{tabular}{>{\raggedright\arraybackslash}p{0.36\linewidth}>{\raggedright\arraybackslash}p{0.22\linewidth}>{\raggedright\arraybackslash}p{0.32\linewidth}}
\toprule
Quantity & scalar model & full-wave field\\
\midrule
Dark ring: radius, height & \SI{0.44}{\milli\meter}, \SI{3.6}{\milli\meter} & 0.44--\SI{0.49}{\milli\meter}, 3.4--\SI{3.7}{\milli\meter}\\
Linking and winding numbers & 1.000--1.002 & 1.000--1.001\\
Half-plane counts equal to one & 120 of 120 & 120 of 120\\
Order of the designed ring (morph) & 1 & 1\\
Overlap & 0.79 with the target & 0.88 with the scalar model\\
Modal $|N_S|$; $|Q_W|$ & 0.998--0.999; 0.995--0.997 & 1.005--1.008; 1.016--1.031\\
\bottomrule
\end{tabular}
\end{table}

\begin{table}[!htbp]
\centering
\caption{\textbf{Full-wave field at three frequencies.}}
\label{tab:freq}
\small
\begin{tabular}{lccc}
\toprule
& \SI{0.975}{\tera\hertz} & \SI{1.000}{\tera\hertz} & \SI{1.025}{\tera\hertz}\\
\midrule
Closed loops (ring and three preimages) & 4 & 4 & 3\\
Linking and winding numbers & 1.000--1.001 & 1.000--1.001 & 0.991--1.004\\
Modal $|N_S|$ & 0.998 & 1.005--1.008 & 1.012--1.025\\
Modal $|Q_W|$ & 0.995--0.996 & 1.016--1.031 & 1.019--1.038\\
Overlap with the scalar model & 0.96 & 0.88 & 0.69\\
Power in $E_z$ & 4.0\% & 4.9\% & 6.5\%\\
Transmitted / reflected power & -- & 65\% / 15\% & --\\
Transmission of the $x$ / $y$ channel & -- & 0.656 / 0.658 & --\\
\bottomrule
\end{tabular}
\end{table}

\sfig{SI_scalar_maps.pdf}{1}{\textbf{Scalar model of the $28\times28$ first-order device.} \textbf{a},~Phase maps $\phi_x$ and $\phi_y$ from equation~(9) of the main text at the cell centers. \textbf{b},\,\textbf{c},~Normalized $|E_x|^2$ and $|E_y|^2$ in the plane $y=0$. \textbf{d},~Preimage loops: the axis ($E_y=0$), the ring ($E_x=0$) and the states $\tilde A/\tilde B=8,\,8\mathrm{i},\,-8$. \textbf{e},~Transverse texture at $z=1.5$, 3 and \SI{4.5}{\milli\meter}: hue, $\arg(E_y/E_x)$; brightness, transverse intensity relative to the plane maximum.}{fig:mapsc}
\sfig{SI_fullwave_maps.pdf}{1}{\textbf{Full-wave field of the complete device at \SI{1}{\tera\hertz}.} \textbf{a},~Phases realized by the selected cells, $\phi_x$ and $\phi_y$. \textbf{b},\,\textbf{c},~Normalized $|E_x|^2$ and $|E_y|^2$ in the plane $y=0$. \textbf{d},~Preimage loops traced in the full-wave field; legend as in Supplementary Fig.~\ref{fig:mapsc}d. \textbf{e},~Transverse texture at $z=1.5$, 3 and \SI{4.5}{\milli\meter}, displayed as in Supplementary Fig.~\ref{fig:mapsc}e.}{fig:mapfw}

\section{Higher-order designs}
\label{sn:general}
This note gives the details of the designs of Fig.~5 and Table~1 of the main text, built from the ring components of Supplementary Table~\ref{tab:designs}.

\paragraph{Ideal fields.} The Skyrme number of equation~(5) of the main text was evaluated over all space with both components scaled to equal power in the focal plane (Supplementary Table~\ref{tab:gen_ideal}, Supplementary Fig.~\ref{fig:orders}a). For every design inside conditions (i)--(iii), $|N_S|$ rises monotonically towards $n\ell$ as the grid is refined, with a residual proportional to the squared grid step; the residual grows with the order because the outer rings lie where the grid, uniform in $\tan^{-1}\xi$, is coarser. In the boundary case $\ell=2n$ the values converge to 0.342, close to the exact non-integer value $\tfrac13$ of Step~13 for two components of equal power; this grid resolves the boundary of space poorly in the boundary case and gives a value about 3\% higher.

\begin{table}[!htbp]
\centering
\caption{\textbf{Skyrme number $|N_S|$ of the ideal fields over all space.} Grids of $N^3$ points with $\xi_{x,y}=\tan(\pi a/2)$; both components scaled to equal power in the focal plane. Last column: extrapolation $q(h)=q_0+ch^2$ from the three finest grids.}
\label{tab:gen_ideal}
\begin{tabular}{lccccccc}
\toprule
Ring component & $\ell$ & $n\ell$ & $N=101$ & $N=201$ & $N=301$ & $N=401$ & $h\to0$\\
\midrule
$n=1$, $(\tfrac34,\tfrac14)$ & 1 & 1 & 0.994 & 0.998 & 0.999 & 1.000 & 1.000\\
$n=1$, $(\tfrac34,\tfrac14)$ & 2 & $\ell=2n$ & 0.340 & 0.341 & 0.342 & 0.342 & 0.342\\
$n=2$, $(3,2,2)$ & 1 & 2 & 1.965 & 1.991 & 1.996 & 1.998 & 2.000\\
$n=2$, $(3,2,2)$ & 2 & 4 & 3.958 & 3.989 & 3.995 & 3.997 & 4.000\\
$n=2$, $(3,2,2)$ & 3 & 6 & 5.932 & 5.983 & 5.992 & 5.996 & 6.000\\
$n=3$, $(4,3,2,2)$ & 1 & 3 & 2.777 & 2.937 & 2.971 & 2.984 & 2.999\\
$n=3$, $(4,3,2,2)$ & 2 & 6 & 5.744 & 5.932 & 5.969 & 5.983 & 5.999\\
$n=3$, $(4,3,2,2)$ & 3 & 9 & 8.728 & 8.929 & 8.968 & 8.982 & 9.000\\
$n=3$, $(4,3,2,2)$ & 5 & 15 & 14.336 & 14.820 & 14.918 & 14.954 & 14.998\\
\bottomrule
\end{tabular}
\end{table}

\paragraph{Sheets and intensity floors.} Each design was realized on the sheet of the device with equation~(9) of the main text and propagated with the scalar model. Counting the dark rings of $E_x$ above an intensity floor gives orders that depend on the floor (Supplementary Table~\ref{tab:gen_sheet}). Low floors admit further rings in the periphery of the beam, and high floors drop dim designed rings; for the three-ring masks with $\ell=1$ and 2 the outer designed ring is dim because $|E_y|^2\propto\rho^{2\ell}e^{-2\rho^2}$ is small there for small $\ell$, although the ring of $E_x$ is the same for every $\ell$.

\begin{table}[!htbp]
\centering
\caption{\textbf{Sheets of $28\times28$ cells, scalar model.} Rings: radius and height of the dark rings of $E_x$ at the 0.3\% floor and the winding of $\arg E_y$ along each. Order: realized order at floors of 0.2, 0.3, 0.5 and 1\%. $I_{\min}$: intensity at the dimmest ring, in percent of the peak of the power-balanced field. $|Q_W|$: regularized modal estimate at $\varepsilon=0.003$ and 0.01. $F$: field overlap with the target.}
\label{tab:gen_sheet}
\resizebox{\textwidth}{!}{\begin{tabular}{ccllccc}
\toprule
$(n,\ell)$ & $n\ell$ & rings: $r$, $z$ (mm), winding & order & $I_{\min}$ (\%) & $|Q_W|$ & $F$\\
\midrule
$(1,1)$ & 1 & 0.44, 3.59, 1 & 1 / 1 / 1 / 1 & 8.89 & 1.00, 1.00 & 0.79\\
$(1,2)$ & $\ell=2n$ & 0.44, 3.59, 2;\ 0.86, 4.08, 2 & 6 / 4 / 2 / 2 & 0.43 & 3.84, 1.99 & 0.82\\
$(2,1)$ & 2 & 0.37, 2.89, 1;\ 0.70, 3.35, 1 & 2 / 2 / 2 / 1 & 0.67 & 1.63, 1.03 & 0.88\\
$(2,2)$ & 4 & 0.37, 2.89, 2;\ 0.70, 3.35, 2 & 6 / 4 / 4 / 4 & 1.97 & 4.06, 3.93 & 0.91\\
$(2,3)$ & 6 & 0.37, 2.89, 3;\ 0.70, 3.35, 3 & 9 / 6 / 6 / 6 & 3.13 & 5.96, 5.88 & 0.91\\
$(3,1)$ & 3 & 0.35, 2.75, 1;\ 0.58, 2.53, 1;\ 0.98, 2.17, 1 & 2 / 3 / 2 / 2 & 0.40 & 1.98, 1.98 & 0.86\\
$(3,2)$ & 6 & 0.35, 2.75, 2;\ 0.58, 2.53, 2 & 4 / 4 / 4 / 4 & 4.41 & 4.05, 3.95 & 0.89\\
$(3,3)$ & 9 & 0.35, 2.75, 3;\ 0.58, 2.53, 3;\ 0.89, 2.77, 3 & 9 / 9 / 9 / 6 & 0.92 & 8.69, 6.22 & 0.89\\
\bottomrule
\end{tabular}}
\end{table}

\paragraph{Continuation from the target.} The order is therefore determined without any floor (main-text Methods). Along the straight path from the target to the realized field, every ring of $E_x$ is traced and followed from step to step (Supplementary Table~\ref{tab:cont}). In every design each ring of the target continues without interruption into a ring of the realized field, with the same orientation and with $\ell$ turns of the phase of $E_y$, and the realized order of these rings is exactly $n\ell$. The closed preimages of $\tilde A/\tilde B=30$ and $-30$ that wrap them link exactly $n\ell$ times. All other rings appear along the path in the periphery. Most enter through the outer edge of the analysis volume, and in the three-ring fields a pair of opposite orientation is created in the periphery, which adds nothing to the order. For the designs of main-text Fig.~5 every designed ring is brighter than every other ring; for the three-ring masks with $\ell=1$ and 2 the outer designed ring is dimmer than the created pair, so that no intensity floor separates them. These are genuine zeros, found by both tracers, and the count over the whole analysis volume, which includes them, is the same for the two tracers.

\begin{table}[!htbp]
\centering
\caption{\textbf{Rings of the $28\times28$ sheets followed from the target} (scalar model, no intensity floor). Designed: rings that continue the rings of the target, and their order; $Lk$, linking of the closed preimages of $\tilde A/\tilde B=\pm30$ that wrap them. Other rings: rings that appear along the path, the order of all rings in the analysis volume ($r\le3.5w_0$) by the half-plane and the grid tracer, and the intensity of the brightest other ring and of the dimmest designed ring, relative to the peak of the power-balanced field.}
\label{tab:cont}
\small
\begin{tabular}{ccccccccc}
\toprule
& & \multicolumn{3}{c}{designed rings} & \multicolumn{2}{c}{all rings} & \multicolumn{2}{c}{intensity (\%)}\\
$(n,\ell)$ & $n\ell$ & rings & order & $Lk$ & order & grid & brightest other & dimmest designed\\
\midrule
$(1,1)$ & 1 & 1 & 1 & 1.000 & 4 & 4 & 0.14 & 8.89\\
$(2,1)$ & 2 & 2 & 2 & 2.000 & 4 & 4 & 0.09 & 0.67\\
$(2,2)$ & 4 & 2 & 4 & 4.000 & 8 & 8 & 0.28 & 1.97\\
$(2,3)$ & 6 & 2 & 6 & 6.001 & 12 & 12 & 0.25 & 3.13\\
$(3,1)$ & 3 & 3 & 3 & 3.000 & 5 & 5 & 0.40 & 0.08\\
$(3,2)$ & 6 & 3 & 6 & 6.000 & 10 & 10 & 0.18 & 0.05\\
$(3,3)$ & 9 & 3 & 9 & 9.000 & 15 & 15 & 0.17 & 0.92\\
\bottomrule
\end{tabular}
\end{table}

\sfig{SI_orders.pdf}{1}{\textbf{Higher-order designs.} \textbf{a},~$|N_S|/n\ell$ of the ideal fields over all space versus the squared grid step. \textbf{b},~Realized order of the $28\times28$ sheets (traced rings and windings) at four intensity floors, for all designs; color, ratio to $n\ell$. \textbf{c},~$|E_x|^2$ of the three-ring sheets in the plane $y=0$ on a logarithmic scale, with the crossings of the traced dark rings (circles).}{fig:orders}

\paragraph{Brightness of the rings.} The ring positions follow from the coefficients alone, and they decide whether a design is practical. Simple choices such as $F(x)\propto(1+x)^n$ satisfy conditions (i)--(iii) but push the outer ring far into the dim periphery of the beam; the coefficients $(3,2,2)$ and $(4,3,2,2)$ pull the rings inwards (Supplementary Table~\ref{tab:designs}). Within a single Gaussian envelope every further ring costs roughly an order of magnitude in the intensity of the dimmest ring.

\section{Range of validity}
\label{sn:range}
The scalar model was used to test how far the design can be pushed away from the device, varying one parameter at a time around it ($\ell=1$, $\kappa=\tfrac12$, $z_0=10\lambda$, $w_0=1.5\lambda$, $28\times28$ cells, illumination waist $6\lambda$), and how much error the cells tolerate (Supplementary Table~\ref{tab:range}). Two limits stand out. The sheet must lie more than about one Rayleigh range before the focus, because closer to the focus $B$ is nearly real and changes sign across the ring, an amplitude zero that a phase-only sheet cannot reproduce. And the illuminated area must match the footprint of the target, because a phase-only cell transmits the full illumination amplitude; a large sheet illuminated far beyond that footprint adds strongly focused light that the paraxial target does not contain. Within these limits the linking holds for every tested design, and random or fabrication errors change the waveform long before they unlink the loops.

\begin{table}[!htbp]
\centering
\caption{\textbf{Range of validity} (scalar model, first-order device unless stated). $Lk$, range of the linking numbers of the traced loops; $F$, overlap with the target or with the error-free device.}
\label{tab:range}
\small
\begin{tabular}{>{\raggedright\arraybackslash}p{0.27\linewidth}>{\raggedright\arraybackslash}p{0.30\linewidth}>{\raggedright\arraybackslash}p{0.35\linewidth}}
\toprule
Parameter & range & result\\
\midrule
Ring parameter $\kappa$ & 0.2--1.2 & $Lk=1.000$--1.003\\
Focal distance, waist & $z_0=7.5\lambda$--$30\lambda$, $w_0=0.75\lambda$--$2\lambda$ & $Lk=0.996$--1.004; $F=0.40$--0.97\\
& $\zeta\le0.71$ & no closed ring\\
Sheet size & 20--64 cells & $Lk=1.000$--1.003\\
& $100\times100$, waist $6\lambda$ & $Lk=1.000$--1.001\\
& $100\times100$, illumination scaled with the sheet & extra zeros cross the ring; unlinked at $z_0=20\lambda$\\
Random cell phase errors & up to \ang{30} r.m.s. & 6/6 linked at every level; $F$ down to 0.92\\
& \ang{45}, \ang{60}, \ang{90} r.m.s. & 5/6, 1/6, 0/6 linked\\
Fabrication deviations & up to $\pm\SI{1.5}{\micro\meter}$ & all 18 realizations linked; $F\ge0.97$\\
\bottomrule
\end{tabular}
\end{table}